\documentclass[pra,letterpaper,aps,10pt,superscriptaddress,twocolumn,floatfix,showpacs]{revtex4-1}
\usepackage{amsmath,amsfonts,dsfont,graphicx,amssymb,microtype,braket,xcolor,upgreek,mathtools,physics}
\usepackage{placeins}
\usepackage[colorlinks, linkcolor=blue, citecolor=blue, urlcolor=blue, breaklinks=true]{hyperref}
\DeclareGraphicsExtensions{.pdf}
\hypersetup{
    colorlinks=true,
    linkcolor=blue,
    citecolor=blue,
    urlcolor=blue,
}

\newcommand{\s}{\sigma}

\newcommand{\td}{\mathrm{d}}

\newcommand{\ts}[1]{_\text{#1}}

\newcommand{\calH}{\mathcal{H}}

\newcommand{\me}{\mathrm{e}}
\newcommand{\mi}{\mathrm{i}}
\usepackage{physics}

\newcommand{\revise}[1]{{\color{black}#1}}
\newcommand{\reviseTwo}[1]{{\color{black}#1}}

\begin{document}

\title{Nonlinear semiclassical spectroscopy of ultrafast molecular polariton dynamics}
\author{Michael Reitz}
\affiliation{Department of Chemistry and Biochemistry, University of California San Diego, La Jolla, California 92093, USA}
\author{Arghadip Koner}
\affiliation{Department of Chemistry and Biochemistry, University of California San Diego, La Jolla, California 92093, USA}
\author{Joel Yuen-Zhou}
\email{joelyuen@ucsd.edu}
\affiliation{Department of Chemistry and Biochemistry, University of California San Diego, La Jolla, California 92093, USA}
\date{\today}

\begin{abstract}
We introduce a theoretical framework that allows for the systematic and efficient  description of the ultrafast nonlinear response of molecular polaritons, i.e., hybrid light-matter states, in the collective regime of large numbers of molecules $\mathcal N$ coupled to the cavity photon mode. Our approach is based on a semiclassical, mean-field evolution of the molecular Hamiltonian and the cavity field, which is complemented by a perturbative expansion of both light and matter counterparts in the input fields entering the cavity. In addition, expansion in terms of the pulse phases enables us to disentangle different excitation pathways in Liouville space, thereby distinguishing contributions to the nonlinear response.  The formalism extends traditional free-space nonlinear spectroscopy by incorporating the feedback of matter onto the light field via the induced polarization. We demonstrate the utility of the framework by applying it to the calculation of pump-probe polariton spectra and show how, by storing the pulses, the cavity facilitates additional excitation pathways, \revise{which can be used to isolate purely bright state contributions}. Our method, which does not scale with $\mathcal N$, is broadly applicable and can be extended to \revise{model} a wide range of current experiments investigating the dynamical nonlinear response of hybrid light-matter states.
\end{abstract}

\maketitle

\begin{figure*}[t]
  \centering
  \includegraphics[width=0.95\textwidth]{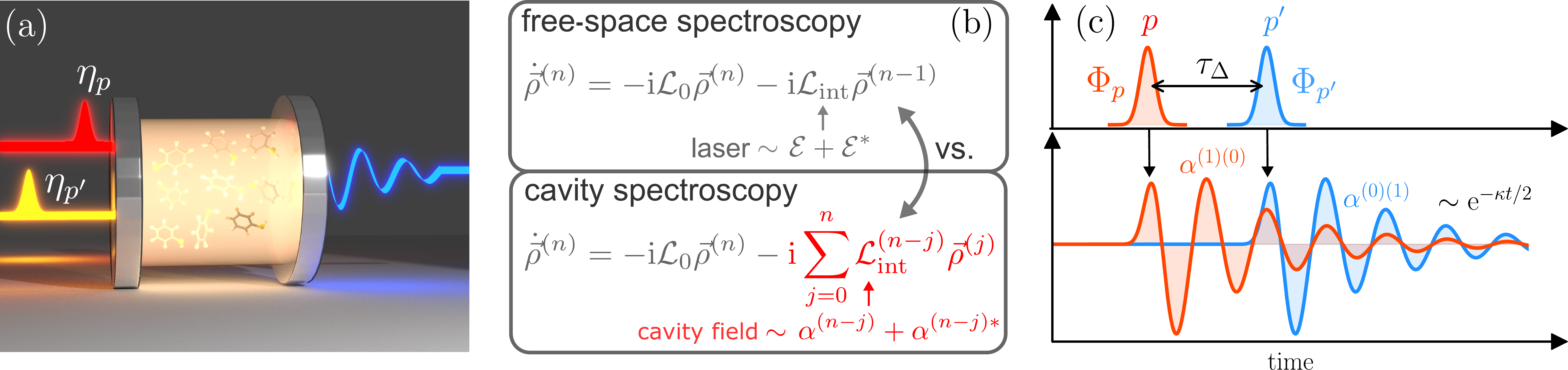}
    \caption{(a) \textit{Schematics.} An optical cavity containing an ensemble of $\mathcal N$ molecules is driven by input pulses with amplitudes $\eta_p$ (pump) and $\eta_{p'}$ (probe). (b) Key distinction between conventional, free-space nonlinear spectroscopy and nonlinear cavity spectroscopy, concerning the evolution of the molecular density matrix. For cavity spectroscopy, the feedback of the molecular polarization onto the cavity field leads to a coupling of the $n$th order molecular density matrix $\vec\rho^{(n)}$ to all lower orders in the cavity field, as opposed to the free-space case, where the interaction is always linear in the laser field that drives the molecules. (c) Pump and probe pulses entering the cavity at time delay $\tau_\Delta$ get stored and prolonged in the cavity on a time scale given by the cavity lifetime $\kappa^{-1}$, thereby allowing the \revise{(linear)} cavity field created by the probe ($\alpha^{(0)(1)}$) to act before the pump ($\alpha^{(1)(0)}$). Additional phases $\Phi_{p,p'}$ imprinted on the pulses can be utilized to separate excitation pathways in Liouville space.}
    \label{fig1}
\end{figure*}

\noindent \textbf{Introduction.}\textemdash Molecular polaritons, quasiparticles resulting from the strong coupling between confined photonic modes and vibrational or electronic excitations of molecules, e.g., inside Fabry-P{\'e}rot cavities, have gained significant attention over recent years, largely due to their potential applications in areas such as polariton chemistry \cite{hutchinson2012modifying, thomas2019tilting}, energy transport \cite{coles2014polariton, zhong2017energy}, condensation and lasing \cite{kenacohen2010room,plumhof2014room}, and the manipulation of optical nonlinearities~\cite{barachati2018tunable, xiang2019manipulating, wang2021large, cheng2022molecular}. Recent advances in electronic \cite{takemura2015dephasing, takemura2015two, delpo2020polariton, fassioli2021femtosecond, mewes2020energy, son2022energy, wu2022optical, russo2024direct, michail2024addressing} and vibrational \cite{xiang2018twodimensional, xiang2019manipulating, xiong2023molecular, sufrin2024probing, dunkelberger2022vibration, duan2021isolating, simpkins2023comment} nonlinear polariton spectroscopy have provided new insight into the nonlinear response of molecular polaritons by revealing the dynamics of the complex interactions and energy relaxation pathways. However, significant challenges and open questions remain in understanding the nonlinear polariton response, particularly regarding the short-time dynamical behavior, i.e., before the cavity photons, and thus the polariton states, have decayed ($\sim$ \reviseTwo{fs-ps} in experiments). The development of an efficient formalism that can address these questions is crucial for advancing the understanding of polariton dynamics across different time scales as observed in pump-probe experiments.

While the theoretical formalism of nonlinear spectroscopy involving the interaction between laser light and matter in free space is well-established  \cite{mukamel1995principles, boyd2008nonlinear, yuenzhou2014ultrafast, jonas2003two,cho2008coherent, gelin2009efficient}, recent works have aimed at constructing a framework for the theory of nonlinear electronic and vibrational polariton spectroscopy~\cite{saurabh2016two, ribeiro2018theory, zhang2023multidimensional, mondal2023quantum, gallego2024coherent, schnappinger2024disentangling}. Some studies have considered direct excitation  of the molecules through the side of the cavity \cite{zhang2023multidimensional}, whereas other works have taken a fully quantum mechanical approach, thereby restricting the scope of the treatment to either long times after light and matter have decoupled, or the explicit simulation of only a few molecules \cite{ribeiro2018theory, gallego2024coherent}.

In this Letter, we introduce a systematic formalism to nonlinear cavity spectroscopy in the limit of large numbers of molecules $\mathcal N$ per photon mode. Our approach is based on a mean-field evolution of the coupled light-matter system \cite{fowler2022efficient, fowlerwrightthesis}, related to semiclassical Maxwell-Liouville approaches which have been developed, e.g., in semiconductor optics and nanophotonics \cite{jahnke1996excitonic, scully1997quantum, lopota2009multiscale, sindelka2010derivation, sukharev2011numerical,li2018mixed, chen2019ehrenfest, bhat2001optical, bhat2006hamiltonian}.  We perturbatively expand both cavity field and density matrix in the input fields driving the cavity and show that the formalism can be regarded as an extension of the conventional free-space perturbative nonlinear spectroscopy framework by Mukamel \cite{mukamel1995principles}, now incorporating nonlinear contributions to the field resulting from the feedback of the molecules onto the cavity dynamics. \revise{Expansion of the nonlinear response in terms of phase components allows us to separate different excitation pathways in Liouville space and identify purely bright state contributions}, using the widely-adopted toolkit of double-sided Feynman diagrams.\\

\noindent \textbf{Mean-field approach.}\textemdash We consider an ensemble of $\mathcal N$ molecules collectively coupled to the single confined mode of an optical cavity at frequency $\omega_c$ [see schematics in Fig.~\ref{fig1}(a)]. Each molecule $j$ is described by an (in principle arbitrary) Hamiltonian $\calH_0^j$ with a corresponding dipole operator $\hat\mu_j$. The interaction between the molecules and the cavity field is described by a Dicke-type Hamiltonian such that the total Hamiltonian describing the full quantum evolution of the coupled light-matter system expresses as (we set $\hbar\equiv 1$)
\begin{align}
\mathcal{H}= \omega_c \hat{a}^\dagger\hat{a}+\sum_{j=1}^\mathcal{N} \calH_0^j +  E_0 (\hat a+\hat a^\dagger) \sum_{j=1}^\mathcal{N} \hat\mu_j ,
\end{align}
with the cavity zero-point amplitude $E_0=\sqrt{\omega_c/(2\epsilon_0\mathcal{V})}$, where $\mathcal{V}$ is the cavity mode volume and $\epsilon_0$ is the vacuum permittivity.
To facilitate the understanding of the formalism, we initially assume that the cavity field is driven by only a single input field with amplitude $\eta$ at a central carrier frequency $\omega_\ell$ and with a Gaussian pulse shape described by the temporal envelope function $f(t)$ with a temporal pulse width $\tau_w$.

 The limit of large $\mathcal N$ allows us to apply  a mean-field approximation (exact for $\mathcal N\to \infty$), which amounts to setting $\expval{\hat a}=\alpha (t)$ \cite{fowler2022efficient}. Additionally, for illustration, we take all molecules to be identical, i.e., \revise{$\calH_0^j=\calH_0$}, $\hat\mu_j=\hat\mu$, although extension to disordered ensembles offers no conceptual difficulty. The simplified mean-field Hamiltonian then expresses as
\begin{align}
\label{eq:meanfieldham}
\calH\ts{MF}=\calH_0 +E_0 \left(\alpha (t)+\alpha^* (t)\right)\hat{\mu}.
\end{align}
The dynamical evolution of the \revise{(now single-molecule)} density matrix $\rho$ is governed by the master equation $\partial_t\rho=\mi[\rho,\mathcal{H}\ts{MF}]+\mathcal{D}[\rho]$. Here, the dissipator $\mathcal{D}[\rho]$ represents possible additional loss channels of the molecule (e.g., dephasing at rate $\gamma_\phi$, non-radiative decay,...) that may be added, e.g., in Lindblad form. However, a complete description of the mean-field problem also involves solving the classical equation for the cavity field amplitude which undergoes damping at rate $\kappa$ (photon loss)
\begin{align}
\label{eq:cavityfield}
\dot\alpha (t)&=-\left(\frac{\kappa}{2}+\mi\omega_c\right)\alpha (t)-\mi\mathcal{N}E_0 P(t)-\eta f(t) \me^{-\mi\omega_\ell t},
\end{align}
containing the induced molecular polarization $P(t)=\mathrm{Tr}[\hat\mu\rho (t)]$ of all $\mathcal N$ molecules. Altogether, this describes a self-consistent, semiclassical evolution of the coupled light-matter system, where the matter component evolves quantum mechanically, while the part describing the electromagnetic field undergoes classical evolution and is driven by both the input field as well as by the collective molecular polarization feeding back into the cavity field. \revise{While Eq.~\eqref{eq:cavityfield} assumes a simplified, single-mode description of the electromagnetic field, we discuss the straightforward extension to multimode cavities in the Supplementary Material (SM) \cite{sm}.} \\

\noindent \textbf{Perturbative expansion.}\textemdash  While the above set of mean-field equations can be readily solved numerically, there is generally no closed analytical solution due to the nonlinearity of the molecular transitions. To gain more insight into the structure of the nonlinear response, we therefore proceed, akin to standard nonlinear spectroscopy, with a perturbative expansion of both cavity field \revise{\textit{and}} matter density matrix in terms of the input field amplitude, i.e.,
\begin{align}
\alpha (t) = \sum_{n=0}^\infty \eta^n \alpha^{(n)}(t), \quad \rho (t)=\sum_{n=0}^\infty \eta^n \rho^{(n)}(t).
\end{align}
From this, also other perturbative quantities can be derived such as, e.g., the $n$th order polarization $P^{(n)} (t)=\mathrm{Tr}[\hat \mu \rho^{(n)}(t)]$  \footnote{Note that the $\rho^{(n)}$ are not true density matrices, i.e., they do not obey the properties of density matrices such as the conservation of the trace}.

\begin{figure}[t]
    \centering
    \includegraphics[width=0.8\columnwidth]{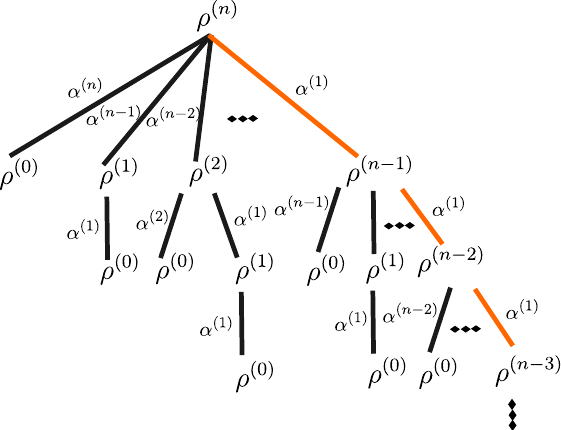}
    \caption{\textit{Density matrix evolution.} Graph illustrating the possible integrals contributing to the dynamical evolution of the $n$th order \revise{molecular} density matrix $\rho^{(n)}(t)$. The orange path on the right illustrates the integral that is obtained for conventional free-space spectroscopy, corresponding to a sequence of first-order interactions linear in the input. The feedback of the nonlinear polarization onto the cavity field adds $2^{n-1}-1$ other possible pathways.}
    \label{fig2}
\end{figure}

Furthermore, for both numerical and analytical considerations, it is advantageous to convert the master equation for the density matrix onto matrix-vector form, i.e., map the density matrix $\rho$ onto a vector  $\vec\rho $ and the superoperator acting on the density matrix onto a matrix  (also referred to as Liouville space) [see SM \cite{sm} for details on vectorization] \cite{mukamel1995principles, horn1994topics, amshallem2015approaches}.
Then, one obtains a closed system for the cavity field amplitude and the vectorized density matrix up to $n$th order
\begin{subequations}
\label{eq:perturbativeeqs}
 \begin{align}
 \label{eq:nfield}
\dot\alpha^{(n)}&=-\left(\frac{\kappa}{2}+\mi\omega_c\right)\alpha^{(n)}-\mi\mathcal{N}E_0 P^{(n)}-\delta_{1n}f(t)\me^{-\mi\omega_\ell t},\\
 \label{eq:ndensity}
\dot{\vec\rho}^{(n)}&=-\mi\mathcal{L}_0\vec{\rho}^{(n)}-\mi E_0\sum_{j=0}^n \left(\alpha^{(n-j)}+\alpha^{(n-j)*}\right)\mathcal{L}_\mu\vec{\rho}^{(j)},
\end{align}
\end{subequations}
in terms of the vectorized commutators $[\mathcal{H}_0, \rho^{(n)}]\to \mathcal{L}_0\vec{\rho}^{(n)}$ and $[\hat\mu, \rho^{(j)}]\to\mathcal{L}_\mu \vec{\rho}^{(j)}$, where $\mathcal{L}_0$, $\mathcal{L}_\mu$ are the free molecular and interaction Liouvillians, respectively.
In particular, Eqs.~\eqref{eq:perturbativeeqs} indicate that only the first-order cavity field is driven directly by the input \revise{pulse} while at higher orders, the molecular density matrix is driven by lower orders of the cavity field. Importantly, Eq.~\eqref{eq:ndensity} shows that the (cavity) field driving the molecular density matrix can have nonlinear contributions in the input field, as opposed to free-space spectroscopy where the interaction Liouvillian driving the evolution of the density matrix is always linear in the (laser) input field [see Fig.~\ref{fig1}(b)] \cite{mukamel1995principles}. In other words, the electric field experienced by the molecules is not the laser field interacting with the cavity from outside, but a distorted version of it due to the evolving material polarization inside the cavity. While for conventional spectroscopy with well-separated pulses, only a single multidimensional integral is obtained as solution for the $n$th order density matrix, many additional integrals arise in the cavity scenario ($2^{n-1}-1$, see diagram in Fig.~\ref{fig2}). Also, a straightforward analysis of Eqs.~\eqref{eq:perturbativeeqs} reveals that the first order ($n=1$) recovers the standard linear polariton response, characterized by a Rabi splitting proportional to $\sqrt{\mathcal{N}}$, arising from the hybridization of the cavity with the collective molecular bright mode~\cite{larocca1998biexcitons, agranovic2003cavity, litinskaya2004fast, gonzalezballestero2016uncoupled, camposgonzalez2022generalization, koner2024linear}. \\

\begin{figure}[b]
    \centering
    \includegraphics[width=1.0\columnwidth]{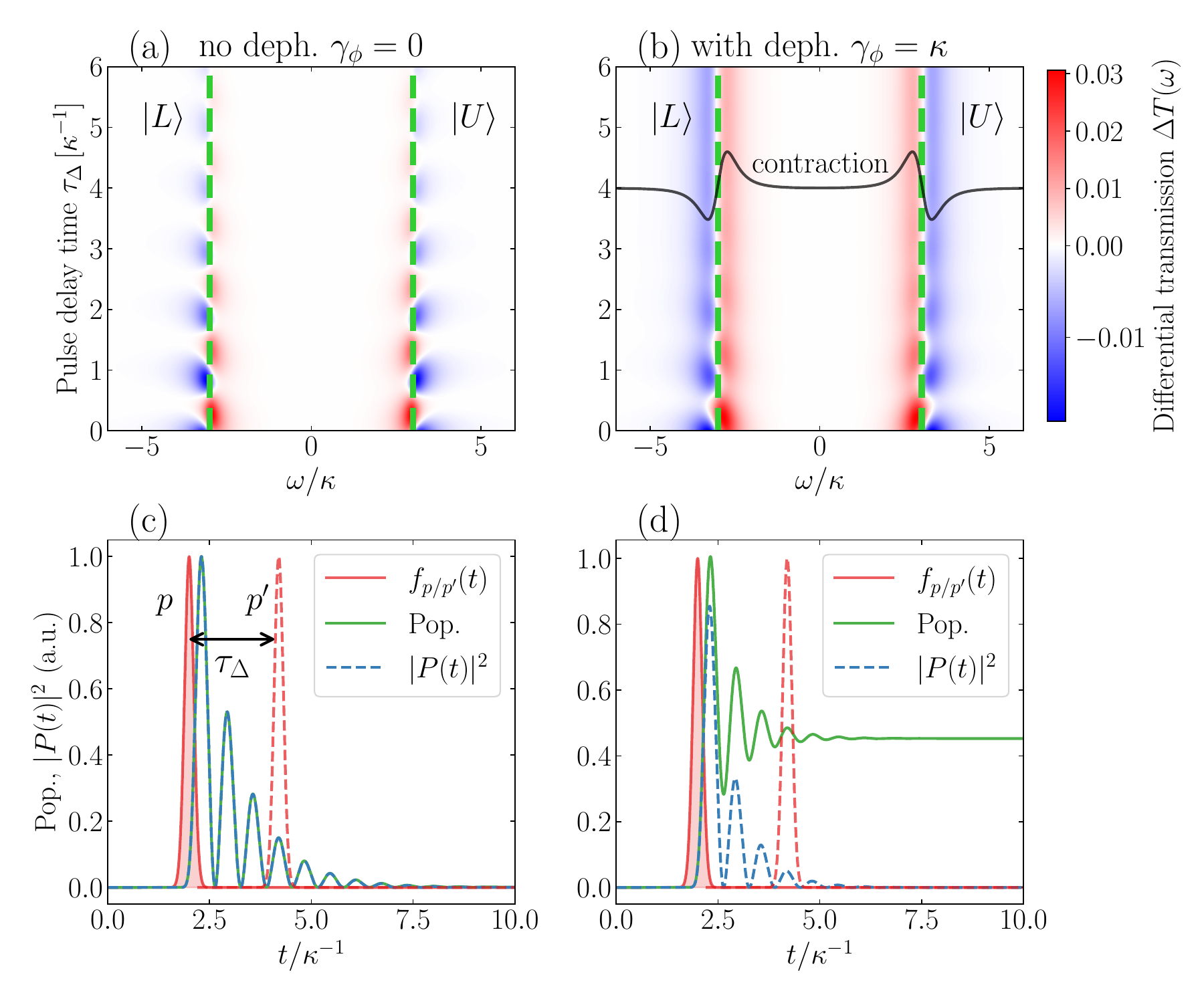}
    \caption{\textit{Differential transmission.} DT spectra $\Delta T (\omega)$ for a 2LS with resonance frequency $\omega_0$ as a function of the delay time between the pump and probe pulse $\tau_\Delta$ for (a) weak ($\gamma_\phi =0$), and (b) strong ($\gamma_\phi=\kappa$) dephasing of the transition. The green dashed lines show the location of the (linear response) polariton frequencies. The black curve in (b) shows a cross section through the DT at $\tau_\Delta=4\kappa^{-1}$. \revise{The results are plotted in a frame rotating at the central pulse frequency.} \revise{Panels (c) and (d) depict} the dynamics of the population and polarization $|P(t)|^2$ created by the initial pump pulse arriving at $\tau_p=2\kappa^{-1}$ which are then probed by the second pulse at delay $\tau_\Delta$. The 2LS is initialized in the ground state, with parameters $g\sqrt{\mathcal{N}}=3\kappa$, $\tau_w=0.05\kappa^{-1}$, $\omega_0=\omega_c$, and \revise{carrier} frequency $\omega_{p}=\omega_{p'}=\omega_0$ for both pulses.}
    \label{fig3}
\end{figure}

\noindent \textbf{Differential transmission.}\textemdash We now consider two pulses driving the cavity (pump $p$ and probe $p'$), with input fields proportional to $ \eta_j f_j(t-\tau_j)\me^{-\mi\omega_j t}$ where $\eta_j$, $f_j$, $\omega_j$, $\tau_j$ for $j\in\{p,p'\}$ denote the amplitude, envelope, carrier frequency, and arrival time of the pump and probe pulses, respectively. We assume a time delay between pump and probe of $\tau_\Delta=\tau_{p'}-\tau_p\geq 0$ [see sketch in Fig.~\ref{fig1}(c)].  We proceed with a perturbative expansion in both pump ($p$) and probe ($p'$) amplitudes (this can be generalized to arbitrary many pulses)
\begin{subequations}
\begin{align}
\alpha (t)& = \sum_{n,m=0}^\infty\eta_p^n\eta_{p'}^m\alpha^{(n)(m)} (t), \\
 \rho(t)&=\sum_{n,m=0}^\infty\eta_p^n\eta_{p'}^m \rho^{(n)(m)}(t),
\end{align}
\end{subequations}
where the first and second indices refer to the orders in the pump and probe fields, respectively. This allows us to derive a perturbative set of equations for $\alpha^{(n)(m)} $, $\vec{\rho}^{(n)(m)}$ similar to Eqs.~\eqref{eq:perturbativeeqs}, now describing pump-probe spectroscopy  (see SM~\cite{sm} for details).

\begin{figure*}[t]
    \centering
    \includegraphics[width=0.97\textwidth]{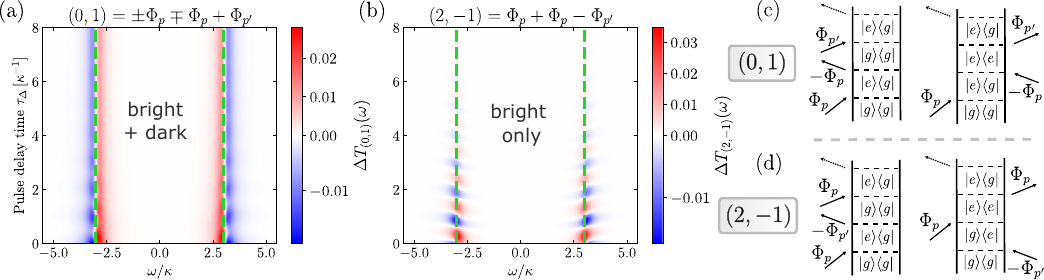}
    \caption{\textit{Separating phase contributions to the nonlinear polariton response.}  (a), (b) Plot of phase contributions $(0,1)$ and $(2,-1)$ to $\Delta T (\omega)$ as a function of the delay time between the pulses. The $(2,-1)$-contribution decays on a  time scale of \revise{$\sim (\kappa+\gamma_\phi)^{-1}$, as it relies on the probe acting on the coherence created by the pump, therefore probing the bright state only.} The green dashed lines show the (linear response) polariton frequencies. We chose $g\sqrt{\mathcal{N}}=3\kappa$, $\gamma_\phi=0.1\kappa$, $\omega_{p}=\omega_{p'}=\omega_0$, and a temporal width of $\tau_w=0.1\kappa^{-1}$ for both pulses. The double-sided Feynman diagrams in (c) and (d) represent processes contributing to the different phase combinations of \revise{the third-order cavity field} $\alpha^{(2)(1)}$. The dotted arrows indicate the decay of the coherence.}
    \label{fig4}
\end{figure*}

From the solution for the total cavity field, the cavity transmission and reflection can be obtained using standard input-output theory \cite{steck2007quantum} \revise{(valid as long as the system is not in the ultrastrong coupling regime~\cite{ciuti2006input, friskkockum2019ultrastrong})}. Since nonlinearities are typically weak, instead of absolute quantities, differential (nonlinear) corrections to the response are commonly considered.
For the collinear geometry in question, we define the differential transmission (DT) spectrum by subtracting the (normalized) probe transmission without the pump from the probe transmission in the presence of the pump \revise{field}
\begin{align}
\Delta T (\omega)=T_{p'}^\text{pump\, on}(\omega)-T_{p'}^\text{pump\, off}(\omega),
\end{align}
where the transmission is related to the intracavity field as $T_{p'}(\omega)=|(\kappa/2)\alpha (\omega)/[\eta_{p'}f_{p'}(\omega)]|^2$. The contribution $T_{p'}^\text{pump\, on}(\omega)$ depends on the time delay between the pulses $\tau_\Delta$. By considering only the lowest (third) order correction due to pump and probe, the DT may be expressed as  (see SM for derivation~\cite{sm})
\begin{align}
\label{eq:dt}
\Delta T (\omega)\approx\left(\frac{\kappa}{2}\right)^2\eta_p^2\frac{2\mathrm{Re}\left[\alpha^{(0)(1)*}(\omega)\alpha^{(2)(1)}(\omega)\right]}{f_{p'}(\omega)^2},
\end{align}
\reviseTwo{describing interference [four-wave mixing (4WM)] between the probe (acting as local oscillator) and the third-order cavity field $\alpha^{(2)(1)}$}. Beware that Eq.~\eqref{eq:dt} is not the usual pump-probe signal; instead, it contains different phase contributions (see paragraph below). The DT is plotted as a function of delay time in Figs.~\ref{fig3}(a), (b) for two-level systems (2LSs) with transition dipole moment $\mu_{ge}$ coupled to the cavity at strength $g=\mu_{ge} E_0$, for two different degrees of dephasing. The plots below in Figs.~\ref{fig3}(c), (d) show the dynamics of the polarization and population created by an initial pump pulse which are then probed after a time $\tau_\Delta$. At early times, oscillatory behavior around the polariton frequencies is observed, corresponding to coherences between upper and lower polariton. Dephasing leads to a \revise{faster} decay of coherence and the creation of stationary population in the molecules, even after the cavity photon has decayed. In the long-time limit, this manifests as a contracted Rabi splitting, characterized by a DT as shown by the black cross-section curve in Fig.~\ref{fig3}(b) \cite{xiang2018twodimensional, dunkelberger2022vibration, ribeiro2018theory}.\\

\noindent \textbf{Separating excitation pathways.}\textemdash  The nonlinear signal stemming from the $n$th order density matrix generally contains many possible excitation pathways. For instance, even for free-space spectroscopy of a 2LS, there are eight Feynman diagrams contributing to the third-order nonlinear response \cite{mukamel1995principles}. In experiments, it is therefore highly desired to isolate certain excitation pathways and thereby understand the different contributions to the nonlinear signal, which here, owing to the collinear geometry in question, can be achieved by phase cycling. \revise{This method is equivalent to the more standard phase matching procedure with oblique pulses, which, however obfuscates our formalism as it requires the description of a multimode cavity \cite{yuenzhou2014ultrafast, jonas2003two, hybl1998two, tian2003femtosecond, tan2008theory, yang2020controlling}}.

 To this end, we will consider the effect of  additional phases applied to the pump ($ \Phi_p$) and probe ($ \Phi_{p'}$)  pulses.  These phases are subsequently also imprinted onto the cavity field and the molecular density matrix, enabling us to disentangle different excitation pathways characterized by specific phase sequences~\cite{mukamel1995principles, yuenzhou2014ultrafast}.  The cavity field and density matrix are now expanded as
\begin{subequations}
\begin{align}
\alpha (t) &=\sum_{n,m, \vec{v}}\eta_p^n \eta_{p'}^m \me^{-\mi\vec{v}\cdot \vec{\Phi}} \alpha^{(n)(m)}_{\vec{v}} (t),\\
 \rho (t) &=\sum_{n,m, \vec{v}}\eta_p^n \eta_{p'}^m \me^{-\mi\vec{v}\cdot \vec{\Phi}}\rho^{(n)(m)}_{\vec{v}}(t),
\end{align}
\end{subequations}
where  $\vec\Phi =(\Phi_p,\Phi_{p'})^\top$ denotes the vector of pump and probe phases, and $\vec v=(v_p , v_{p'})^\top$ with $v_p, v_{p'}\in\mathbb{Z}_0$ denotes the phase coefficients which also depend on the pump and probe orders $v_p (n)$, $v_{p'} (m)$ (see SM \cite{sm}). The raw DT signal from Eq.~\eqref{eq:dt} can then be separated into different phase contributions 
\begin{align}
\label{eq:deltav}
\Delta T_{\vec{v}}(\omega)\approx\left(\frac{\kappa}{2}\right)^2\eta_p^2\frac{2\mathrm{Re}\left[\alpha^{(0)(1)*}_{(0,-1)}(\omega)\alpha^{(2)(1)}_{\vec{v}}(\omega)\right]}{f_{p'}(\omega)^2},
\end{align}
\reviseTwo{where $(0,-1)$ is the only phase contribution for the conjugate probe field \cite{sm} and the total signal is obtained as $\Delta T=\sum_{\vec{v}}\Delta T_{\vec{v}}$.}
\revise{Two phase combinations contribute to $\alpha^{(2)(1)}$: $(0,1)$ and $(2,-1)$ [along with the complex conjugates $(0,-1)$ and $(-2,1)$], corresponding to pump--probe and double-quantum coherence (DQC) spectroscopies, respectively [Figs.~4(a,b)]. The pump--probe pathway $(0,1)\equiv \pm\Phi_p\!\mp\!\Phi_p\!+\! \Phi_{p'}$ creates populations and zeroth-order quantum coherences via two pump pulses of opposite phase [Fig.~\ref{fig4}(c)]. The DQC pathway $(2,-1)\equiv 2\Phi_p\!-\!\Phi_{p'}$ is negligible outside the cavity because it requires the two-level system to be excited twice. However, inside the cavity, the probe can act on the pump-induced coherence within the lifetime of the bright state $(\kappa+\gamma_\phi)^{-1}$ [left diagram in Fig.~\ref{fig4}(d)], enabling this pathway to arise within this time interval. The right diagram in Fig.~\ref{fig4}(d) does not contribute to the \reviseTwo{probe} DT since the pump has not acted yet. Consequently, the $(2,-1)$ pathway isolates the bright state only, in contrast to the \((0,1)\) pathway, which accesses both bright and dark states (and probes only dephased dark states in the long time limit). \reviseTwo{Eq.~\eqref{eq:deltav} satisfies the phase matching conditions for degenerate 4WM with two pulses \cite{yuenzhou2014ultrafast}.}  } 
\\

\noindent \textbf{Discussion and conclusions.}\textemdash We have presented an efficient and broadly applicable framework for nonlinear polariton spectroscopy, based on a mean-field evolution of the coupled light-matter system. The formalism extends conventional free-space spectroscopy by accounting for the feedback of the  molecular polarization onto the cavity field and allows for an understanding of the polariton response beyond the stationary limit \cite{ribeiro2018theory}. Expanding the cavity field and molecular density matrix in terms of the input fields' amplitudes and phases \revise{provides} a systematic understanding of the various contribution to the nonlinear polariton response, \revise{and allows to separate purely bright state contributions from dark reservoir states}.

The formalism offers potential for extension and generalization via multiple avenues. For instance, while our current treatment relies on a semiclassical mean-field description, valuable extensions could involve capturing nonclassical photon statistics \revise{or many-body correlations among particles}, e.g., by using cumulant \revise{or cluster} expansion techniques \cite{fowlerwright2023determining, axt1994dynamics}.
Other areas of interest are the study of spatially-resolved nonlinear polariton spectroscopy, which would involve generalizing the zero-dimensional cavity Eq.~\eqref{eq:cavityfield} to the 3D set of cavity modes afforded by classical electromagnetic simulations \cite{gruetzmacher2003finite, zhou2024simulating,yang2023enabling, balasubrahmaniyam2023from, xu2023ultrafast}, or cavities exhibiting inherent nonlinearities \cite{finkelstein2023non}.
Furthermore, while we have focused on pump-probe spectroscopy involving two pulses, the formalism can be readily applied to 2D or multidimensional coherent spectroscopy, \revise{which may also help to disentangle pure electronic dephasing from vibronic relaxation channels}.
Finally, let us emphasize that our approach is not limited to molecular polaritons, but could also be extended to other strongly coupled light-matter platforms, such as exciton-polaritons in inorganic semiconductor microcavities \cite{fryett2018cavity} or atomic \revise{ensembles}. \\
%%%%%%%%%%%%%%%%%%%%%%%%%%%%%%%%%%%%%%%%%%%%%%%%%%%%%%%%%%%%%%%%%%%%%%%%%%%%%%%%%%%%%%%%%%%%%%%%%%%%%

%%%%%%%%%%%%%%%%%%%%%%%%%%%%%%%%%%%%%%%%%%%%%%%%%%%%%%%%%%%%%%%%%%%%%%%%%%%%%%%%%%%%%%%%%%%%%%%%%%%%%

\noindent \textbf{Acknowledgments.}\textemdash We are grateful to Wei Xiong for providing valuable feedback and comments on the manuscript. This research was primarily supported by the Air Force Office
of Scientific Research (AFOSR) through the Multi-University Research Initiative (MURI)
program no.~FA9550-22-1-0317. \\

\bibliographystyle{apsrev4-1-custom}
\bibliography{NonlinearRefs}

\begin{thebibliography}{77}%
\makeatletter
\providecommand \@ifxundefined [1]{%
 \@ifx{#1\undefined}
}%
\providecommand \@ifnum [1]{%
 \ifnum #1\expandafter \@firstoftwo
 \else \expandafter \@secondoftwo
 \fi
}%
\providecommand \@ifx [1]{%
 \ifx #1\expandafter \@firstoftwo
 \else \expandafter \@secondoftwo
 \fi
}%
\providecommand \natexlab [1]{#1}%
\providecommand \enquote  [1]{``#1''}%
\providecommand \bibnamefont  [1]{#1}%
\providecommand \bibfnamefont [1]{#1}%
\providecommand \citenamefont [1]{#1}%
\providecommand \href@noop [0]{\@secondoftwo}%
\providecommand \href [0]{\begingroup \@sanitize@url \@href}%
\providecommand \@href[1]{\@@startlink{#1}\@@href}%
\providecommand \@@href[1]{\endgroup#1\@@endlink}%
\providecommand \@sanitize@url [0]{\catcode `\\12\catcode `\$12\catcode
  `\&12\catcode `\#12\catcode `\^12\catcode `\_12\catcode `\%12\relax}%
\providecommand \@@startlink[1]{}%
\providecommand \@@endlink[0]{}%
\providecommand \url  [0]{\begingroup\@sanitize@url \@url }%
\providecommand \@url [1]{\endgroup\@href {#1}{\urlprefix }}%
\providecommand \urlprefix  [0]{URL }%
\providecommand \Eprint [0]{\href }%
\providecommand \doibase [0]{http://dx.doi.org/}%
\providecommand \selectlanguage [0]{\@gobble}%
\providecommand \bibinfo  [0]{\@secondoftwo}%
\providecommand \bibfield  [0]{\@secondoftwo}%
\providecommand \translation [1]{[#1]}%
\providecommand \BibitemOpen [0]{}%
\providecommand \bibitemStop [0]{}%
\providecommand \bibitemNoStop [0]{.\EOS\space}%
\providecommand \EOS [0]{\spacefactor3000\relax}%
\providecommand \BibitemShut  [1]{\csname bibitem#1\endcsname}%
\let\auto@bib@innerbib\@empty
%</preamble>
\bibitem [{\citenamefont {Hutchison} \emph {et~al.}(2012)\citenamefont
  {Hutchison}, \citenamefont {Schwartz}, \citenamefont {Genet}, \citenamefont
  {Devaux}, and \citenamefont {Ebbesen}}]{hutchinson2012modifying}%
  \BibitemOpen
  \bibfield  {author} {\bibinfo {author} {\bibfnamefont {J.~A.} \bibnamefont
  {Hutchison}}, \bibinfo {author} {\bibfnamefont {T.}~\bibnamefont {Schwartz}},
  \bibinfo {author} {\bibfnamefont {C.}~\bibnamefont {Genet}}, \bibinfo
  {author} {\bibfnamefont {E.}~\bibnamefont {Devaux}},  and \bibinfo {author}
  {\bibfnamefont {T.~W.} \bibnamefont {Ebbesen}}, }\bibfield  {title} {\enquote
  {\bibinfo {title} {Modifying chemical landscapes by coupling to vacuum
  fields},} }\href {\doibase https://doi.org/10.1002/anie.201107033} {\bibfield
   {journal} {\bibinfo  {journal} {Angew. Chem. Int. Ed.} }\textbf {\bibinfo
  {volume} {51}}, \bibinfo {pages} {1592--1596} (\bibinfo {year}
  {2012})}\BibitemShut {NoStop}%
\bibitem [{\citenamefont {Thomas} \emph {et~al.}(2019)\citenamefont {Thomas},
  \citenamefont {Lethuillier-Karl}, \citenamefont {Nagarajan}, \citenamefont
  {Vergauwe}, \citenamefont {George}, \citenamefont {Chervy}, \citenamefont
  {Shalabney}, \citenamefont {Devaux}, \citenamefont {Genet}, \citenamefont
  {Moran}, and \citenamefont {Ebbesen}}]{thomas2019tilting}%
  \BibitemOpen
  \bibfield  {author} {\bibinfo {author} {\bibfnamefont {A.}~\bibnamefont
  {Thomas}}, \bibinfo {author} {\bibfnamefont {L.}~\bibnamefont
  {Lethuillier-Karl}}, \bibinfo {author} {\bibfnamefont {K.}~\bibnamefont
  {Nagarajan}}, \bibinfo {author} {\bibfnamefont {R.~M.~A.} \bibnamefont
  {Vergauwe}}, \bibinfo {author} {\bibfnamefont {J.}~\bibnamefont {George}},
  \bibinfo {author} {\bibfnamefont {T.}~\bibnamefont {Chervy}}, \bibinfo
  {author} {\bibfnamefont {A.}~\bibnamefont {Shalabney}}, \bibinfo {author}
  {\bibfnamefont {E.}~\bibnamefont {Devaux}}, \bibinfo {author} {\bibfnamefont
  {C.}~\bibnamefont {Genet}}, \bibinfo {author} {\bibfnamefont
  {J.}~\bibnamefont {Moran}},  and \bibinfo {author} {\bibfnamefont {T.~W.}
  \bibnamefont {Ebbesen}}, }\bibfield  {title} {\enquote {\bibinfo {title}
  {Tilting a ground-state reactivity landscape by vibrational strong
  coupling},} }\href {\doibase 10.1126/science.aau7742} {\bibfield  {journal}
  {\bibinfo  {journal} {Science} }\textbf {\bibinfo {volume} {363}}, \bibinfo
  {pages} {615--619} (\bibinfo {year} {2019})}\BibitemShut {NoStop}%
\bibitem [{\citenamefont {Coles} \emph {et~al.}(2014)\citenamefont {Coles},
  \citenamefont {Somaschi}, \citenamefont {Michetti}, \citenamefont {Clark},
  \citenamefont {Lagoudakis}, \citenamefont {Savvidis}, and \citenamefont
  {Lidzey}}]{coles2014polariton}%
  \BibitemOpen
  \bibfield  {author} {\bibinfo {author} {\bibfnamefont {D.~M.} \bibnamefont
  {Coles}}, \bibinfo {author} {\bibfnamefont {N.}~\bibnamefont {Somaschi}},
  \bibinfo {author} {\bibfnamefont {P.}~\bibnamefont {Michetti}}, \bibinfo
  {author} {\bibfnamefont {C.}~\bibnamefont {Clark}}, \bibinfo {author}
  {\bibfnamefont {P.~G.} \bibnamefont {Lagoudakis}}, \bibinfo {author}
  {\bibfnamefont {P.~G.} \bibnamefont {Savvidis}},  and \bibinfo {author}
  {\bibfnamefont {D.~G.} \bibnamefont {Lidzey}}, }\bibfield  {title} {\enquote
  {\bibinfo {title} {Polariton-mediated energy transfer between organic dyes in
  a strongly coupled optical microcavity},} }\href {\doibase 10.1038/nmat3950}
  {\bibfield  {journal} {\bibinfo  {journal} {Nat. Mater.} }\textbf {\bibinfo
  {volume} {13}}, \bibinfo {pages} {712--719} (\bibinfo {year}
  {2014})}\BibitemShut {NoStop}%
\bibitem [{\citenamefont {Zhong} \emph {et~al.}(2017)\citenamefont {Zhong},
  \citenamefont {Chervy}, \citenamefont {Zhang}, \citenamefont {Thomas},
  \citenamefont {George}, \citenamefont {Genet}, \citenamefont {Hutchison}, and
  \citenamefont {Ebbesen}}]{zhong2017energy}%
  \BibitemOpen
  \bibfield  {author} {\bibinfo {author} {\bibfnamefont {X.}~\bibnamefont
  {Zhong}}, \bibinfo {author} {\bibfnamefont {T.}~\bibnamefont {Chervy}},
  \bibinfo {author} {\bibfnamefont {L.}~\bibnamefont {Zhang}}, \bibinfo
  {author} {\bibfnamefont {A.}~\bibnamefont {Thomas}}, \bibinfo {author}
  {\bibfnamefont {J.}~\bibnamefont {George}}, \bibinfo {author} {\bibfnamefont
  {C.}~\bibnamefont {Genet}}, \bibinfo {author} {\bibfnamefont {J.~A.}
  \bibnamefont {Hutchison}},  and \bibinfo {author} {\bibfnamefont {T.~W.}
  \bibnamefont {Ebbesen}}, }\bibfield  {title} {\enquote {\bibinfo {title}
  {Energy transfer between spatially separated entangled molecules},} }\href
  {\doibase https://doi.org/10.1002/anie.201703539} {\bibfield  {journal}
  {\bibinfo  {journal} {Angew. Chem. Int. Ed.} }\textbf {\bibinfo {volume}
  {56}}, \bibinfo {pages} {9034--9038} (\bibinfo {year} {2017})}\BibitemShut
  {NoStop}%
\bibitem [{\citenamefont {K{\'e}na-Cohen} and \citenamefont
  {Forrest}(2010)}]{kenacohen2010room}%
  \BibitemOpen
  \bibfield  {author} {\bibinfo {author} {\bibfnamefont {S.}~\bibnamefont
  {K{\'e}na-Cohen}} and \bibinfo {author} {\bibfnamefont {S.~R.} \bibnamefont
  {Forrest}}, }\bibfield  {title} {\enquote {\bibinfo {title} {Room-temperature
  polariton lasing in an organic single-crystal microcavity},} }\href {\doibase
  10.1038/nphoton.2010.86} {\bibfield  {journal} {\bibinfo  {journal} {Nat.
  Phot.} }\textbf {\bibinfo {volume} {4}}, \bibinfo {pages} {371--375}
  (\bibinfo {year} {2010})}\BibitemShut {NoStop}%
\bibitem [{\citenamefont {Plumhof} \emph {et~al.}(2014)\citenamefont {Plumhof},
  \citenamefont {St{\"o}ferle}, \citenamefont {Mai}, \citenamefont {Scherf},
  and \citenamefont {Mahrt}}]{plumhof2014room}%
  \BibitemOpen
  \bibfield  {author} {\bibinfo {author} {\bibfnamefont {J.~D.} \bibnamefont
  {Plumhof}}, \bibinfo {author} {\bibfnamefont {T.}~\bibnamefont
  {St{\"o}ferle}}, \bibinfo {author} {\bibfnamefont {L.}~\bibnamefont {Mai}},
  \bibinfo {author} {\bibfnamefont {U.}~\bibnamefont {Scherf}},  and \bibinfo
  {author} {\bibfnamefont {R.~F.} \bibnamefont {Mahrt}}, }\bibfield  {title}
  {\enquote {\bibinfo {title} {Room-temperature {B}ose--{E}instein condensation
  of cavity exciton--polaritons in a polymer},} }\href {\doibase
  10.1038/nmat3825} {\bibfield  {journal} {\bibinfo  {journal} {Nat. Mater.}
  }\textbf {\bibinfo {volume} {13}}, \bibinfo {pages} {247--252} (\bibinfo
  {year} {2014})}\BibitemShut {NoStop}%
\bibitem [{\citenamefont {Barachati} \emph {et~al.}(2018)\citenamefont
  {Barachati}, \citenamefont {Simon}, \citenamefont {Getmanenko}, \citenamefont
  {Barlow}, \citenamefont {Marder}, and \citenamefont
  {K{\'e}na-Cohen}}]{barachati2018tunable}%
  \BibitemOpen
  \bibfield  {author} {\bibinfo {author} {\bibfnamefont {F.}~\bibnamefont
  {Barachati}}, \bibinfo {author} {\bibfnamefont {J.}~\bibnamefont {Simon}},
  \bibinfo {author} {\bibfnamefont {Y.~A.} \bibnamefont {Getmanenko}}, \bibinfo
  {author} {\bibfnamefont {S.}~\bibnamefont {Barlow}}, \bibinfo {author}
  {\bibfnamefont {S.~R.} \bibnamefont {Marder}},  and \bibinfo {author}
  {\bibfnamefont {S.}~\bibnamefont {K{\'e}na-Cohen}}, }\bibfield  {title}
  {\enquote {\bibinfo {title} {Tunable third-harmonic generation from
  polaritons in the ultrastrong coupling regime},} }\href {\doibase
  10.1021/acsphotonics.7b00305} {\bibfield  {journal} {\bibinfo  {journal} {ACS
  Photonics} }\textbf {\bibinfo {volume} {5}}, \bibinfo {pages} {119--125}
  (\bibinfo {year} {2018})}\BibitemShut {NoStop}%
\bibitem [{\citenamefont {Xiang} \emph {et~al.}(2019)\citenamefont {Xiang},
  \citenamefont {Ribeiro}, \citenamefont {Li}, \citenamefont {Dunkelberger},
  \citenamefont {Simpkins}, \citenamefont {Yuen-Zhou}, and \citenamefont
  {Xiong}}]{xiang2019manipulating}%
  \BibitemOpen
  \bibfield  {author} {\bibinfo {author} {\bibfnamefont {B.}~\bibnamefont
  {Xiang}}, \bibinfo {author} {\bibfnamefont {R.~F.} \bibnamefont {Ribeiro}},
  \bibinfo {author} {\bibfnamefont {Y.}~\bibnamefont {Li}}, \bibinfo {author}
  {\bibfnamefont {A.~D.} \bibnamefont {Dunkelberger}}, \bibinfo {author}
  {\bibfnamefont {B.~B.} \bibnamefont {Simpkins}}, \bibinfo {author}
  {\bibfnamefont {J.}~\bibnamefont {Yuen-Zhou}},  and \bibinfo {author}
  {\bibfnamefont {W.}~\bibnamefont {Xiong}}, }\bibfield  {title} {\enquote
  {\bibinfo {title} {Manipulating optical nonlinearities of molecular
  polaritons by delocalization},} }\href {\doibase 10.1126/sciadv.aax5196}
  {\bibfield  {journal} {\bibinfo  {journal} {Sci. Adv.} }\textbf {\bibinfo
  {volume} {5}}, \bibinfo {pages} {eaax5196} (\bibinfo {year}
  {2019})}\BibitemShut {NoStop}%
\bibitem [{\citenamefont {Wang} \emph {et~al.}(2021)\citenamefont {Wang},
  \citenamefont {Seidel}, \citenamefont {Nagarajan}, \citenamefont {Chervy},
  \citenamefont {Genet}, and \citenamefont {Ebbesen}}]{wang2021large}%
  \BibitemOpen
  \bibfield  {author} {\bibinfo {author} {\bibfnamefont {K.}~\bibnamefont
  {Wang}}, \bibinfo {author} {\bibfnamefont {M.}~\bibnamefont {Seidel}},
  \bibinfo {author} {\bibfnamefont {K.}~\bibnamefont {Nagarajan}}, \bibinfo
  {author} {\bibfnamefont {T.}~\bibnamefont {Chervy}}, \bibinfo {author}
  {\bibfnamefont {C.}~\bibnamefont {Genet}},  and \bibinfo {author}
  {\bibfnamefont {T.}~\bibnamefont {Ebbesen}}, }\bibfield  {title} {\enquote
  {\bibinfo {title} {Large optical nonlinearity enhancement under electronic
  strong coupling},} }\href {\doibase 10.1038/s41467-021-21739-7} {\bibfield
  {journal} {\bibinfo  {journal} {Nat. Commun.} }\textbf {\bibinfo {volume}
  {12}}, \bibinfo {pages} {1486} (\bibinfo {year} {2021})}\BibitemShut
  {NoStop}%
\bibitem [{\citenamefont {Cheng} \emph {et~al.}(2022)\citenamefont {Cheng},
  \citenamefont {Krainova}, \citenamefont {Brigeman}, \citenamefont {Khanna},
  \citenamefont {Shedge}, \citenamefont {Isborn}, \citenamefont {Yuen-Zhou},
  and \citenamefont {Giebink}}]{cheng2022molecular}%
  \BibitemOpen
  \bibfield  {author} {\bibinfo {author} {\bibfnamefont {C.-Y.} \bibnamefont
  {Cheng}}, \bibinfo {author} {\bibfnamefont {N.}~\bibnamefont {Krainova}},
  \bibinfo {author} {\bibfnamefont {A.~N.} \bibnamefont {Brigeman}}, \bibinfo
  {author} {\bibfnamefont {A.}~\bibnamefont {Khanna}}, \bibinfo {author}
  {\bibfnamefont {S.}~\bibnamefont {Shedge}}, \bibinfo {author} {\bibfnamefont
  {C.}~\bibnamefont {Isborn}}, \bibinfo {author} {\bibfnamefont
  {J.}~\bibnamefont {Yuen-Zhou}},  and \bibinfo {author} {\bibfnamefont {N.~C.}
  \bibnamefont {Giebink}}, }\bibfield  {title} {\enquote {\bibinfo {title}
  {Molecular polariton electroabsorption},} }\href {\doibase
  10.1038/s41467-022-35589-4} {\bibfield  {journal} {\bibinfo  {journal} {Nat.
  Commun.} }\textbf {\bibinfo {volume} {13}}, \bibinfo {pages} {7937} (\bibinfo
  {year} {2022})}\BibitemShut {NoStop}%
\bibitem [{\citenamefont {Takemura} \emph
  {et~al.}(2015{\natexlab{a}})\citenamefont {Takemura}, \citenamefont
  {Anderson}, \citenamefont {Trebaol}, \citenamefont {Biswas}, \citenamefont
  {Oberli}, \citenamefont {Portella-Oberli}, and \citenamefont
  {Deveaud}}]{takemura2015dephasing}%
  \BibitemOpen
  \bibfield  {author} {\bibinfo {author} {\bibfnamefont {N.}~\bibnamefont
  {Takemura}}, \bibinfo {author} {\bibfnamefont {M.~D.} \bibnamefont
  {Anderson}}, \bibinfo {author} {\bibfnamefont {S.}~\bibnamefont {Trebaol}},
  \bibinfo {author} {\bibfnamefont {S.}~\bibnamefont {Biswas}}, \bibinfo
  {author} {\bibfnamefont {D.~Y.} \bibnamefont {Oberli}}, \bibinfo {author}
  {\bibfnamefont {M.~T.} \bibnamefont {Portella-Oberli}},  and \bibinfo
  {author} {\bibfnamefont {B.}~\bibnamefont {Deveaud}}, }\bibfield  {title}
  {\enquote {\bibinfo {title} {Dephasing effects on coherent exciton-polaritons
  and the breakdown of the strong coupling regime},} }\href {\doibase
  10.1103/PhysRevB.92.235305} {\bibfield  {journal} {\bibinfo  {journal} {Phys.
  Rev. B} }\textbf {\bibinfo {volume} {92}}, \bibinfo {pages} {235305}
  (\bibinfo {year} {2015}{\natexlab{a}})}\BibitemShut {NoStop}%
\bibitem [{\citenamefont {Takemura} \emph
  {et~al.}(2015{\natexlab{b}})\citenamefont {Takemura}, \citenamefont
  {Trebaol}, \citenamefont {Anderson}, \citenamefont {Kohnle}, \citenamefont
  {L\'eger}, \citenamefont {Oberli}, \citenamefont {Portella-Oberli}, and
  \citenamefont {Deveaud}}]{takemura2015two}%
  \BibitemOpen
  \bibfield  {author} {\bibinfo {author} {\bibfnamefont {N.}~\bibnamefont
  {Takemura}}, \bibinfo {author} {\bibfnamefont {S.}~\bibnamefont {Trebaol}},
  \bibinfo {author} {\bibfnamefont {M.~D.} \bibnamefont {Anderson}}, \bibinfo
  {author} {\bibfnamefont {V.}~\bibnamefont {Kohnle}}, \bibinfo {author}
  {\bibfnamefont {Y.}~\bibnamefont {L\'eger}}, \bibinfo {author} {\bibfnamefont
  {D.~Y.} \bibnamefont {Oberli}}, \bibinfo {author} {\bibfnamefont {M.~T.}
  \bibnamefont {Portella-Oberli}},  and \bibinfo {author} {\bibfnamefont
  {B.}~\bibnamefont {Deveaud}}, }\bibfield  {title} {\enquote {\bibinfo {title}
  {Two-dimensional {F}ourier transform spectroscopy of exciton-polaritons and
  their interactions},} }\href {\doibase 10.1103/PhysRevB.92.125415} {\bibfield
   {journal} {\bibinfo  {journal} {Phys. Rev. B} }\textbf {\bibinfo {volume}
  {92}}, \bibinfo {pages} {125415} (\bibinfo {year}
  {2015}{\natexlab{b}})}\BibitemShut {NoStop}%
\bibitem [{\citenamefont {DelPo} \emph {et~al.}(2020)\citenamefont {DelPo},
  \citenamefont {Kudisch}, \citenamefont {Park}, \citenamefont {Khan},
  \citenamefont {Fassioli}, \citenamefont {Fausti}, \citenamefont {Rand}, and
  \citenamefont {Scholes}}]{delpo2020polariton}%
  \BibitemOpen
  \bibfield  {author} {\bibinfo {author} {\bibfnamefont {C.~A.} \bibnamefont
  {DelPo}}, \bibinfo {author} {\bibfnamefont {B.}~\bibnamefont {Kudisch}},
  \bibinfo {author} {\bibfnamefont {K.~H.} \bibnamefont {Park}}, \bibinfo
  {author} {\bibfnamefont {S.-U.-Z.} \bibnamefont {Khan}}, \bibinfo {author}
  {\bibfnamefont {F.}~\bibnamefont {Fassioli}}, \bibinfo {author}
  {\bibfnamefont {D.}~\bibnamefont {Fausti}}, \bibinfo {author} {\bibfnamefont
  {B.~P.} \bibnamefont {Rand}},  and \bibinfo {author} {\bibfnamefont {G.~D.}
  \bibnamefont {Scholes}}, }\bibfield  {title} {\enquote {\bibinfo {title}
  {Polariton transitions in femtosecond transient absorption studies of
  ultrastrong light--molecule coupling},} }\href {\doibase
  10.1021/acs.jpclett.0c00247} {\bibfield  {journal} {\bibinfo  {journal} {J.
  Phys. Chem. Lett.} }\textbf {\bibinfo {volume} {11}}, \bibinfo {pages}
  {2667--2674} (\bibinfo {year} {2020})}\BibitemShut {NoStop}%
\bibitem [{\citenamefont {Fassioli} \emph {et~al.}(2021)\citenamefont
  {Fassioli}, \citenamefont {Park}, \citenamefont {Bard}, and \citenamefont
  {Scholes}}]{fassioli2021femtosecond}%
  \BibitemOpen
  \bibfield  {author} {\bibinfo {author} {\bibfnamefont {F.}~\bibnamefont
  {Fassioli}}, \bibinfo {author} {\bibfnamefont {K.~H.} \bibnamefont {Park}},
  \bibinfo {author} {\bibfnamefont {S.~E.} \bibnamefont {Bard}},  and \bibinfo
  {author} {\bibfnamefont {G.~D.} \bibnamefont {Scholes}}, }\bibfield  {title}
  {\enquote {\bibinfo {title} {Femtosecond photophysics of molecular
  polaritons},} }\href {\doibase 10.1021/acs.jpclett.1c03183} {\bibfield
  {journal} {\bibinfo  {journal} {J. Phys. Chem. Lett.} }\textbf {\bibinfo
  {volume} {12}}, \bibinfo {pages} {11444--11459} (\bibinfo {year}
  {2021})}\BibitemShut {NoStop}%
\bibitem [{\citenamefont {Mewes} \emph {et~al.}(2020)\citenamefont {Mewes},
  \citenamefont {Wang}, \citenamefont {Ingle}, \citenamefont {B{\"o}rjesson},
  and \citenamefont {Chergui}}]{mewes2020energy}%
  \BibitemOpen
  \bibfield  {author} {\bibinfo {author} {\bibfnamefont {L.}~\bibnamefont
  {Mewes}}, \bibinfo {author} {\bibfnamefont {M.}~\bibnamefont {Wang}},
  \bibinfo {author} {\bibfnamefont {R.~A.} \bibnamefont {Ingle}}, \bibinfo
  {author} {\bibfnamefont {K.}~\bibnamefont {B{\"o}rjesson}},  and \bibinfo
  {author} {\bibfnamefont {M.}~\bibnamefont {Chergui}}, }\bibfield  {title}
  {\enquote {\bibinfo {title} {Energy relaxation pathways between light-matter
  states revealed by coherent two-dimensional spectroscopy},} }\href {\doibase
  10.1038/s42005-020-00424-z} {\bibfield  {journal} {\bibinfo  {journal}
  {Commun. Phys.} }\textbf {\bibinfo {volume} {3}}, \bibinfo {pages} {157}
  (\bibinfo {year} {2020})}\BibitemShut {NoStop}%
\bibitem [{\citenamefont {Son} \emph {et~al.}(2022)\citenamefont {Son},
  \citenamefont {Armstrong}, \citenamefont {Allen}, \citenamefont {Dhavamani},
  \citenamefont {Arnold}, and \citenamefont {Zanni}}]{son2022energy}%
  \BibitemOpen
  \bibfield  {author} {\bibinfo {author} {\bibfnamefont {M.}~\bibnamefont
  {Son}}, \bibinfo {author} {\bibfnamefont {Z.~T.} \bibnamefont {Armstrong}},
  \bibinfo {author} {\bibfnamefont {R.~T.} \bibnamefont {Allen}}, \bibinfo
  {author} {\bibfnamefont {A.}~\bibnamefont {Dhavamani}}, \bibinfo {author}
  {\bibfnamefont {M.~S.} \bibnamefont {Arnold}},  and \bibinfo {author}
  {\bibfnamefont {M.~T.} \bibnamefont {Zanni}}, }\bibfield  {title} {\enquote
  {\bibinfo {title} {Energy cascades in donor-acceptor exciton-polaritons
  observed by ultrafast two-dimensional white-light spectroscopy},} }\href
  {\doibase 10.1038/s41467-022-35046-2} {\bibfield  {journal} {\bibinfo
  {journal} {Nat. Commun.} }\textbf {\bibinfo {volume} {13}}, \bibinfo {pages}
  {7305} (\bibinfo {year} {2022})}\BibitemShut {NoStop}%
\bibitem [{\citenamefont {Wu} \emph {et~al.}(2022)\citenamefont {Wu},
  \citenamefont {Finkelstein-Shapiro}, \citenamefont {Wang}, \citenamefont
  {Rosenkampff}, \citenamefont {Yartsev}, \citenamefont {Pascher},
  \citenamefont {Nguyen-Phan}, \citenamefont {Cogdell}, \citenamefont
  {B{\"o}rjesson}, and \citenamefont {Pullerits}}]{wu2022optical}%
  \BibitemOpen
  \bibfield  {author} {\bibinfo {author} {\bibfnamefont {F.}~\bibnamefont
  {Wu}}, \bibinfo {author} {\bibfnamefont {D.}~\bibnamefont
  {Finkelstein-Shapiro}}, \bibinfo {author} {\bibfnamefont {M.}~\bibnamefont
  {Wang}}, \bibinfo {author} {\bibfnamefont {I.}~\bibnamefont {Rosenkampff}},
  \bibinfo {author} {\bibfnamefont {A.}~\bibnamefont {Yartsev}}, \bibinfo
  {author} {\bibfnamefont {T.}~\bibnamefont {Pascher}}, \bibinfo {author}
  {\bibfnamefont {T.~C.} \bibnamefont {Nguyen-Phan}}, \bibinfo {author}
  {\bibfnamefont {R.}~\bibnamefont {Cogdell}}, \bibinfo {author} {\bibfnamefont
  {K.}~\bibnamefont {B{\"o}rjesson}},  and \bibinfo {author} {\bibfnamefont
  {T.}~\bibnamefont {Pullerits}}, }\bibfield  {title} {\enquote {\bibinfo
  {title} {Optical cavity-mediated exciton dynamics in photosynthetic light
  harvesting 2 complexes},} }\href {\doibase 10.1038/s41467-022-34613-x}
  {\bibfield  {journal} {\bibinfo  {journal} {Nat. Commun.} }\textbf {\bibinfo
  {volume} {13}}, \bibinfo {pages} {6864} (\bibinfo {year} {2022})}\BibitemShut
  {NoStop}%
\bibitem [{\citenamefont {Russo} \emph {et~al.}(2024)\citenamefont {Russo},
  \citenamefont {Georgiou}, \citenamefont {Genco}, \citenamefont {De~Liberato},
  \citenamefont {Cerullo}, \citenamefont {Lidzey}, \citenamefont {Othonos},
  \citenamefont {Maiuri}, and \citenamefont {Virgili}}]{russo2024direct}%
  \BibitemOpen
  \bibfield  {author} {\bibinfo {author} {\bibfnamefont {M.}~\bibnamefont
  {Russo}}, \bibinfo {author} {\bibfnamefont {K.}~\bibnamefont {Georgiou}},
  \bibinfo {author} {\bibfnamefont {A.}~\bibnamefont {Genco}}, \bibinfo
  {author} {\bibfnamefont {S.}~\bibnamefont {De~Liberato}}, \bibinfo {author}
  {\bibfnamefont {G.}~\bibnamefont {Cerullo}}, \bibinfo {author} {\bibfnamefont
  {D.~G.} \bibnamefont {Lidzey}}, \bibinfo {author} {\bibfnamefont
  {A.}~\bibnamefont {Othonos}}, \bibinfo {author} {\bibfnamefont
  {M.}~\bibnamefont {Maiuri}},  and \bibinfo {author} {\bibfnamefont
  {T.}~\bibnamefont {Virgili}}, }\bibfield  {title} {\enquote {\bibinfo {title}
  {Direct evidence of ultrafast energy delocalization between optically
  hybridized {J}-aggregates in a strongly coupled microcavity},} }\href
  {\doibase https://doi.org/10.1002/adom.202470079} {\bibfield  {journal}
  {\bibinfo  {journal} {Adv. Opt. Mat.} }\textbf {\bibinfo {volume} {12}},
  \bibinfo {pages} {2470079} (\bibinfo {year} {2024})}\BibitemShut {NoStop}%
\bibitem [{\citenamefont {Michail} \emph {et~al.}(2024)\citenamefont {Michail},
  \citenamefont {Rashidi}, \citenamefont {Liu}, \citenamefont {He},
  \citenamefont {Menon}, and \citenamefont {Sfeir}}]{michail2024addressing}%
  \BibitemOpen
  \bibfield  {author} {\bibinfo {author} {\bibfnamefont {E.}~\bibnamefont
  {Michail}}, \bibinfo {author} {\bibfnamefont {K.}~\bibnamefont {Rashidi}},
  \bibinfo {author} {\bibfnamefont {B.}~\bibnamefont {Liu}}, \bibinfo {author}
  {\bibfnamefont {G.}~\bibnamefont {He}}, \bibinfo {author} {\bibfnamefont
  {V.~M.} \bibnamefont {Menon}},  and \bibinfo {author} {\bibfnamefont {M.~Y.}
  \bibnamefont {Sfeir}}, }\bibfield  {title} {\enquote {\bibinfo {title}
  {Addressing the dark state problem in strongly coupled organic
  exciton-polariton systems},} }\href {\doibase 10.1021/acs.nanolett.3c02984}
  {\bibfield  {journal} {\bibinfo  {journal} {Nano Lett.} }\textbf {\bibinfo
  {volume} {24}}, \bibinfo {pages} {557--565} (\bibinfo {year}
  {2024})}\BibitemShut {NoStop}%
\bibitem [{\citenamefont {Xiang} \emph {et~al.}(2018)\citenamefont {Xiang},
  \citenamefont {Ribeiro}, \citenamefont {Dunkelberger}, \citenamefont {Wang},
  \citenamefont {Li}, \citenamefont {Simpkins}, \citenamefont {Owrutsky},
  \citenamefont {Yuen-Zhou}, and \citenamefont
  {Xiong}}]{xiang2018twodimensional}%
  \BibitemOpen
  \bibfield  {author} {\bibinfo {author} {\bibfnamefont {B.}~\bibnamefont
  {Xiang}}, \bibinfo {author} {\bibfnamefont {R.~F.} \bibnamefont {Ribeiro}},
  \bibinfo {author} {\bibfnamefont {A.~D.} \bibnamefont {Dunkelberger}},
  \bibinfo {author} {\bibfnamefont {J.}~\bibnamefont {Wang}}, \bibinfo {author}
  {\bibfnamefont {Y.}~\bibnamefont {Li}}, \bibinfo {author} {\bibfnamefont
  {B.~S.} \bibnamefont {Simpkins}}, \bibinfo {author} {\bibfnamefont {J.~C.}
  \bibnamefont {Owrutsky}}, \bibinfo {author} {\bibfnamefont {J.}~\bibnamefont
  {Yuen-Zhou}},  and \bibinfo {author} {\bibfnamefont {W.}~\bibnamefont
  {Xiong}}, }\bibfield  {title} {\enquote {\bibinfo {title} {Two-dimensional
  infrared spectroscopy of vibrational polaritons},} }\href {\doibase
  10.1073/pnas.1722063115} {\bibfield  {journal} {\bibinfo  {journal} {Proc.
  Natl. Acad. Sci. USA} }\textbf {\bibinfo {volume} {115}}, \bibinfo {pages}
  {4845--4850} (\bibinfo {year} {2018})}\BibitemShut {NoStop}%
\bibitem [{\citenamefont {Xiong}(2023)}]{xiong2023molecular}%
  \BibitemOpen
  \bibfield  {author} {\bibinfo {author} {\bibfnamefont {W.}~\bibnamefont
  {Xiong}}, }\bibfield  {title} {\enquote {\bibinfo {title} {Molecular
  vibrational polariton dynamics: What can polaritons do?}} }\href {\doibase
  10.1021/acs.accounts.2c00796} {\bibfield  {journal} {\bibinfo  {journal}
  {Acc. Chem. Res.} }\textbf {\bibinfo {volume} {56}}, \bibinfo {pages}
  {776--786} (\bibinfo {year} {2023})}\BibitemShut {NoStop}%
\bibitem [{\citenamefont {Sufrin} \emph {et~al.}(2024)\citenamefont {Sufrin},
  \citenamefont {Cohn}, and \citenamefont {Chuntonov}}]{sufrin2024probing}%
  \BibitemOpen
  \bibfield  {author} {\bibinfo {author} {\bibfnamefont {S.}~\bibnamefont
  {Sufrin}}, \bibinfo {author} {\bibfnamefont {B.}~\bibnamefont {Cohn}},  and
  \bibinfo {author} {\bibfnamefont {L.}~\bibnamefont {Chuntonov}}, }\bibfield
  {title} {\enquote {\bibinfo {title} {Probing the anharmonicity of vibrational
  polaritons with double-quantum two-dimensional infrared spectroscopy},}
  }\href {https://doi.org/10.1515/nanoph-2023-0683} {\bibfield  {journal}
  {\bibinfo  {journal} {Nanophotonics} }\textbf {\bibinfo {volume} {13}},
  \bibinfo {pages} {2523--2530} (\bibinfo {year} {2024})}\BibitemShut {NoStop}%
\bibitem [{\citenamefont {Dunkelberger} \emph {et~al.}(2022)\citenamefont
  {Dunkelberger}, \citenamefont {Simpkins}, \citenamefont {Vurgaftman}, and
  \citenamefont {Owrutsky}}]{dunkelberger2022vibration}%
  \BibitemOpen
  \bibfield  {author} {\bibinfo {author} {\bibfnamefont {A.~D.} \bibnamefont
  {Dunkelberger}}, \bibinfo {author} {\bibfnamefont {B.~S.} \bibnamefont
  {Simpkins}}, \bibinfo {author} {\bibfnamefont {I.}~\bibnamefont
  {Vurgaftman}},  and \bibinfo {author} {\bibfnamefont {J.~C.} \bibnamefont
  {Owrutsky}}, }\bibfield  {title} {\enquote {\bibinfo {title}
  {Vibration-cavity polariton chemistry and dynamics},} }\href {\doibase
  https://doi.org/10.1146/annurev-physchem-082620-014627} {\bibfield  {journal}
  {\bibinfo  {journal} {Annu. Rev. Phys. Chem.} }\textbf {\bibinfo {volume}
  {73}}, \bibinfo {pages} {429--451} (\bibinfo {year} {2022})}\BibitemShut
  {NoStop}%
\bibitem [{\citenamefont {Duan} \emph {et~al.}(2021)\citenamefont {Duan},
  \citenamefont {Mastron}, \citenamefont {Song}, and \citenamefont
  {Kubarych}}]{duan2021isolating}%
  \BibitemOpen
  \bibfield  {author} {\bibinfo {author} {\bibfnamefont {R.}~\bibnamefont
  {Duan}}, \bibinfo {author} {\bibfnamefont {J.~N.} \bibnamefont {Mastron}},
  \bibinfo {author} {\bibfnamefont {Y.}~\bibnamefont {Song}},  and \bibinfo
  {author} {\bibfnamefont {K.~J.} \bibnamefont {Kubarych}}, }\bibfield  {title}
  {\enquote {\bibinfo {title} {Isolating polaritonic 2{D}-{I}{R} transmission
  spectra},} }\href {\doibase 10.1021/acs.jpclett.1c03198} {\bibfield
  {journal} {\bibinfo  {journal} {J. Chem. Phys. Lett.} }\textbf {\bibinfo
  {volume} {12}}, \bibinfo {pages} {11406--11414} (\bibinfo {year}
  {2021})}\BibitemShut {NoStop}%
\bibitem [{\citenamefont {Simpkins} \emph {et~al.}(2023)\citenamefont
  {Simpkins}, \citenamefont {Yang}, \citenamefont {Dunkelberger}, \citenamefont
  {Vurgaftman}, \citenamefont {Owrutsky}, and \citenamefont
  {Xiong}}]{simpkins2023comment}%
  \BibitemOpen
  \bibfield  {author} {\bibinfo {author} {\bibfnamefont {B.~S.} \bibnamefont
  {Simpkins}}, \bibinfo {author} {\bibfnamefont {Z.}~\bibnamefont {Yang}},
  \bibinfo {author} {\bibfnamefont {A.~D.} \bibnamefont {Dunkelberger}},
  \bibinfo {author} {\bibfnamefont {I.}~\bibnamefont {Vurgaftman}}, \bibinfo
  {author} {\bibfnamefont {J.~C.} \bibnamefont {Owrutsky}},  and \bibinfo
  {author} {\bibfnamefont {W.}~\bibnamefont {Xiong}}, }\bibfield  {title}
  {\enquote {\bibinfo {title} {Comment on ``{I}solating polaritonic 2{D}-{I}{R}
  transmission spectra''},} }\href {\doibase 10.1021/acs.jpclett.2c01264}
  {\bibfield  {journal} {\bibinfo  {journal} {J. Phys. Chem. Lett.} }\textbf
  {\bibinfo {volume} {14}}, \bibinfo {pages} {983--988} (\bibinfo {year}
  {2023})}\BibitemShut {NoStop}%
\bibitem [{\citenamefont {Mukamel}(1995)}]{mukamel1995principles}%
  \BibitemOpen
  \bibfield  {author} {\bibinfo {author} {\bibfnamefont {S.}~\bibnamefont
  {Mukamel}}, }\href {https://books.google.com/books?id=k_7uAAAAMAAJ} {\emph
  {\bibinfo {title} {Principles of Nonlinear Optical Spectroscopy}}}, Oxford
  {S}eries in {O}ptical and {I}maging {S}ciences (\bibinfo  {publisher} {Oxford
  University Press, New York}, \bibinfo {year} {1995})\BibitemShut {NoStop}%
\bibitem [{\citenamefont {Boyd}(2008)}]{boyd2008nonlinear}%
  \BibitemOpen
  \bibfield  {author} {\bibinfo {author} {\bibfnamefont {R.~W.} \bibnamefont
  {Boyd}}, }\href@noop {} {\emph {\bibinfo {title} {Nonlinear Optics, Third
  Edition}}}, \bibinfo {edition} {3rd} ed. (\bibinfo  {publisher} {Academic
  Press, Inc.}, \bibinfo {address} {USA}, \bibinfo {year} {2008})\BibitemShut
  {NoStop}%
\bibitem [{\citenamefont {Yuen-Zhou} \emph {et~al.}(2014)\citenamefont
  {Yuen-Zhou}, \citenamefont {Krich}, \citenamefont {Kassal}, \citenamefont
  {Johnson}, and \citenamefont {Aspuru-Guzik}}]{yuenzhou2014ultrafast}%
  \BibitemOpen
  \bibfield  {author} {\bibinfo {author} {\bibfnamefont {J.}~\bibnamefont
  {Yuen-Zhou}}, \bibinfo {author} {\bibfnamefont {J.~J.} \bibnamefont {Krich}},
  \bibinfo {author} {\bibfnamefont {I.}~\bibnamefont {Kassal}}, \bibinfo
  {author} {\bibfnamefont {A.~S.} \bibnamefont {Johnson}},  and \bibinfo
  {author} {\bibfnamefont {A.}~\bibnamefont {Aspuru-Guzik}}, }\href {\doibase
  10.1088/978-0-750-31062-8} {\emph {\bibinfo {title} {Ultrafast
  Spectroscopy}}}, 2053-2563 (\bibinfo  {publisher} {IOP Publishing}, \bibinfo
  {year} {2014})\BibitemShut {NoStop}%
\bibitem [{\citenamefont {Jonas}(2003)}]{jonas2003two}%
  \BibitemOpen
  \bibfield  {author} {\bibinfo {author} {\bibfnamefont {D.~M.} \bibnamefont
  {Jonas}}, }\bibfield  {title} {\enquote {\bibinfo {title} {Two-dimensional
  femtosecond spectroscopy},} }\href {\doibase
  https://doi.org/10.1146/annurev.physchem.54.011002.103907} {\bibfield
  {journal} {\bibinfo  {journal} {Ann. Rev. Phys. Chem.} }\textbf {\bibinfo
  {volume} {54}}, \bibinfo {pages} {425--463} (\bibinfo {year}
  {2003})}\BibitemShut {NoStop}%
\bibitem [{\citenamefont {Cho}(2008)}]{cho2008coherent}%
  \BibitemOpen
  \bibfield  {author} {\bibinfo {author} {\bibfnamefont {M.}~\bibnamefont
  {Cho}}, }\bibfield  {title} {\enquote {\bibinfo {title} {Coherent
  two-dimensional optical spectroscopy},} }\href {\doibase 10.1021/cr078377b}
  {\bibfield  {journal} {\bibinfo  {journal} {Chem. Rev.} }\textbf {\bibinfo
  {volume} {108}}, \bibinfo {pages} {1331--1418} (\bibinfo {year}
  {2008})}\BibitemShut {NoStop}%
\bibitem [{\citenamefont {Gelin} \emph {et~al.}(2009)\citenamefont {Gelin},
  \citenamefont {Egorova}, and \citenamefont {Domcke}}]{gelin2009efficient}%
  \BibitemOpen
  \bibfield  {author} {\bibinfo {author} {\bibfnamefont {M.~F.} \bibnamefont
  {Gelin}}, \bibinfo {author} {\bibfnamefont {D.}~\bibnamefont {Egorova}},  and
  \bibinfo {author} {\bibfnamefont {W.}~\bibnamefont {Domcke}}, }\bibfield
  {title} {\enquote {\bibinfo {title} {Efficient calculation of time- and
  frequency-resolved four-wave-mixing signals},} }\href {\doibase
  10.1021/ar900045d} {\bibfield  {journal} {\bibinfo  {journal} {Acc. Chem.
  Res.} }\textbf {\bibinfo {volume} {42}}, \bibinfo {pages} {1290--1298}
  (\bibinfo {year} {2009})}\BibitemShut {NoStop}%
\bibitem [{\citenamefont {Saurabh} and \citenamefont
  {Mukamel}(2016)}]{saurabh2016two}%
  \BibitemOpen
  \bibfield  {author} {\bibinfo {author} {\bibfnamefont {P.}~\bibnamefont
  {Saurabh}} and \bibinfo {author} {\bibfnamefont {S.}~\bibnamefont {Mukamel}},
  }\bibfield  {title} {\enquote {\bibinfo {title} {{Two-dimensional infrared
  spectroscopy of vibrational polaritons of molecules in an optical cavity}},}
  }\href {\doibase 10.1063/1.4944492} {\bibfield  {journal} {\bibinfo
  {journal} {J. Chem. Phys.} }\textbf {\bibinfo {volume} {144}}, \bibinfo
  {pages} {124115} (\bibinfo {year} {2016})}\BibitemShut {NoStop}%
\bibitem [{\citenamefont {F.~Ribeiro} \emph {et~al.}(2018)\citenamefont
  {F.~Ribeiro}, \citenamefont {Dunkelberger}, \citenamefont {Xiang},
  \citenamefont {Xiong}, \citenamefont {Simpkins}, \citenamefont {Owrutsky},
  and \citenamefont {Yuen-Zhou}}]{ribeiro2018theory}%
  \BibitemOpen
  \bibfield  {author} {\bibinfo {author} {\bibfnamefont {R.}~\bibnamefont
  {F.~Ribeiro}}, \bibinfo {author} {\bibfnamefont {A.~D.} \bibnamefont
  {Dunkelberger}}, \bibinfo {author} {\bibfnamefont {B.}~\bibnamefont {Xiang}},
  \bibinfo {author} {\bibfnamefont {W.}~\bibnamefont {Xiong}}, \bibinfo
  {author} {\bibfnamefont {B.~S.} \bibnamefont {Simpkins}}, \bibinfo {author}
  {\bibfnamefont {J.~C.} \bibnamefont {Owrutsky}},  and \bibinfo {author}
  {\bibfnamefont {J.}~\bibnamefont {Yuen-Zhou}}, }\bibfield  {title} {\enquote
  {\bibinfo {title} {Theory for nonlinear spectroscopy of vibrational
  polaritons},} }\href {\doibase 10.1021/acs.jpclett.8b01176} {\bibfield
  {journal} {\bibinfo  {journal} {J. Phys. Chem. Lett.} }\textbf {\bibinfo
  {volume} {9}}, \bibinfo {pages} {3766--3771} (\bibinfo {year}
  {2018})}\BibitemShut {NoStop}%
\bibitem [{\citenamefont {Zhang} \emph {et~al.}(2023)\citenamefont {Zhang},
  \citenamefont {Nie}, \citenamefont {Lei}, and \citenamefont
  {Mukamel}}]{zhang2023multidimensional}%
  \BibitemOpen
  \bibfield  {author} {\bibinfo {author} {\bibfnamefont {Z.}~\bibnamefont
  {Zhang}}, \bibinfo {author} {\bibfnamefont {X.}~\bibnamefont {Nie}}, \bibinfo
  {author} {\bibfnamefont {D.}~\bibnamefont {Lei}},  and \bibinfo {author}
  {\bibfnamefont {S.}~\bibnamefont {Mukamel}}, }\bibfield  {title} {\enquote
  {\bibinfo {title} {Multidimensional coherent spectroscopy of molecular
  polaritons: Langevin approach},} }\href {\doibase
  10.1103/PhysRevLett.130.103001} {\bibfield  {journal} {\bibinfo  {journal}
  {Phys. Rev. Lett.} }\textbf {\bibinfo {volume} {130}}, \bibinfo {pages}
  {103001} (\bibinfo {year} {2023})}\BibitemShut {NoStop}%
\bibitem [{\citenamefont {Mondal} \emph {et~al.}(2023)\citenamefont {Mondal},
  \citenamefont {Koessler}, \citenamefont {Provazza}, \citenamefont
  {Vamivakas}, \citenamefont {Cundiff}, \citenamefont {Krauss}, and
  \citenamefont {Huo}}]{mondal2023quantum}%
  \BibitemOpen
  \bibfield  {author} {\bibinfo {author} {\bibfnamefont {M.~E.} \bibnamefont
  {Mondal}}, \bibinfo {author} {\bibfnamefont {E.~R.} \bibnamefont {Koessler}},
  \bibinfo {author} {\bibfnamefont {J.}~\bibnamefont {Provazza}}, \bibinfo
  {author} {\bibfnamefont {A.~N.} \bibnamefont {Vamivakas}}, \bibinfo {author}
  {\bibfnamefont {S.~T.} \bibnamefont {Cundiff}}, \bibinfo {author}
  {\bibfnamefont {T.~D.} \bibnamefont {Krauss}},  and \bibinfo {author}
  {\bibfnamefont {P.}~\bibnamefont {Huo}}, }\bibfield  {title} {\enquote
  {\bibinfo {title} {{Quantum dynamics simulations of the 2D spectroscopy for
  exciton polaritons}},} }\href {\doibase 10.1063/5.0166188} {\bibfield
  {journal} {\bibinfo  {journal} {J. Chem. Phys.} }\textbf {\bibinfo {volume}
  {159}}, \bibinfo {pages} {094102} (\bibinfo {year} {2023})}\BibitemShut
  {NoStop}%
\bibitem [{\citenamefont {Gallego-Valencia} \emph {et~al.}(2024)\citenamefont
  {Gallego-Valencia}, \citenamefont {Mewes}, \citenamefont {Feist}, and
  \citenamefont {Sanz-Vicario}}]{gallego2024coherent}%
  \BibitemOpen
  \bibfield  {author} {\bibinfo {author} {\bibfnamefont {D.}~\bibnamefont
  {Gallego-Valencia}}, \bibinfo {author} {\bibfnamefont {L.}~\bibnamefont
  {Mewes}}, \bibinfo {author} {\bibfnamefont {J.}~\bibnamefont {Feist}},  and
  \bibinfo {author} {\bibfnamefont {J.~L.} \bibnamefont {Sanz-Vicario}},
  }\bibfield  {title} {\enquote {\bibinfo {title} {Coherent multidimensional
  spectroscopy in polariton systems},} }\href {\doibase
  10.1103/PhysRevA.109.063704} {\bibfield  {journal} {\bibinfo  {journal}
  {Phys. Rev. A} }\textbf {\bibinfo {volume} {109}}, \bibinfo {pages} {063704}
  (\bibinfo {year} {2024})}\BibitemShut {NoStop}%
\bibitem [{\citenamefont {Schnappinger} \emph {et~al.}(2024)\citenamefont
  {Schnappinger}, \citenamefont {Falvo}, and \citenamefont
  {Kowalewski}}]{schnappinger2024disentangling}%
  \BibitemOpen
  \bibfield  {author} {\bibinfo {author} {\bibfnamefont {T.}~\bibnamefont
  {Schnappinger}}, \bibinfo {author} {\bibfnamefont {C.}~\bibnamefont {Falvo}},
   and \bibinfo {author} {\bibfnamefont {M.}~\bibnamefont {Kowalewski}},
  }\bibfield  {title} {\enquote {\bibinfo {title} {Disentangling collective
  coupling in vibrational polaritons with double quantum coherence
  spectroscopy},} }\href {\doibase 10.1063/5.0239877} {\bibfield  {journal}
  {\bibinfo  {journal} {J. Chem. Phys.} }\textbf {\bibinfo {volume} {161}},
  \bibinfo {pages} {244107} (\bibinfo {year} {2024})}\BibitemShut {NoStop}%
\bibitem [{\citenamefont {Fowler-Wright} \emph {et~al.}(2022)\citenamefont
  {Fowler-Wright}, \citenamefont {Lovett}, and \citenamefont
  {Keeling}}]{fowler2022efficient}%
  \BibitemOpen
  \bibfield  {author} {\bibinfo {author} {\bibfnamefont {P.}~\bibnamefont
  {Fowler-Wright}}, \bibinfo {author} {\bibfnamefont {B.~W.} \bibnamefont
  {Lovett}},  and \bibinfo {author} {\bibfnamefont {J.}~\bibnamefont
  {Keeling}}, }\bibfield  {title} {\enquote {\bibinfo {title} {Efficient
  many-body non-{M}arkovian dynamics of organic polaritons},} }\href {\doibase
  10.1103/PhysRevLett.129.173001} {\bibfield  {journal} {\bibinfo  {journal}
  {Phys. Rev. Lett.} }\textbf {\bibinfo {volume} {129}}, \bibinfo {pages}
  {173001} (\bibinfo {year} {2022})}\BibitemShut {NoStop}%
\bibitem [{\citenamefont {Fowler-Wright}(2024)}]{fowlerwrightthesis}%
  \BibitemOpen
  \bibfield  {author} {\bibinfo {author} {\bibfnamefont {P.}~\bibnamefont
  {Fowler-Wright}}, }\emph {\bibinfo {title} {Mean-field and cumulant
  approaches to modelling organic polariton physics}}, \href {\doibase
  10.17630/STA/872} {Ph.D. thesis}, \bibinfo  {school} {The University of St
  Andrews} (\bibinfo {year} {2024})\BibitemShut {NoStop}%
\bibitem [{\citenamefont {Jahnke} \emph {et~al.}(1996)\citenamefont {Jahnke},
  \citenamefont {Kira}, \citenamefont {Koch}, \citenamefont {Khitrova},
  \citenamefont {Lindmark}, \citenamefont {Nelson}, \citenamefont {Wick},
  \citenamefont {Berger}, \citenamefont {Lyngnes}, \citenamefont {Gibbs}, and
  \citenamefont {Tai}}]{jahnke1996excitonic}%
  \BibitemOpen
  \bibfield  {author} {\bibinfo {author} {\bibfnamefont {F.}~\bibnamefont
  {Jahnke}}, \bibinfo {author} {\bibfnamefont {M.}~\bibnamefont {Kira}},
  \bibinfo {author} {\bibfnamefont {S.~W.} \bibnamefont {Koch}}, \bibinfo
  {author} {\bibfnamefont {G.}~\bibnamefont {Khitrova}}, \bibinfo {author}
  {\bibfnamefont {E.~K.} \bibnamefont {Lindmark}}, \bibinfo {author}
  {\bibfnamefont {T.~R.} \bibnamefont {Nelson}, \bibfnamefont {Jr.}}, \bibinfo
  {author} {\bibfnamefont {D.~V.} \bibnamefont {Wick}}, \bibinfo {author}
  {\bibfnamefont {J.~D.} \bibnamefont {Berger}}, \bibinfo {author}
  {\bibfnamefont {O.}~\bibnamefont {Lyngnes}}, \bibinfo {author} {\bibfnamefont
  {H.~M.} \bibnamefont {Gibbs}},  and \bibinfo {author} {\bibfnamefont
  {K.}~\bibnamefont {Tai}}, }\bibfield  {title} {\enquote {\bibinfo {title}
  {Excitonic nonlinearities of semiconductor microcavities in the
  nonperturbative regime},} }\href {\doibase 10.1103/PhysRevLett.77.5257}
  {\bibfield  {journal} {\bibinfo  {journal} {Phys. Rev. Lett.} }\textbf
  {\bibinfo {volume} {77}}, \bibinfo {pages} {5257--5260} (\bibinfo {year}
  {1996})}\BibitemShut {NoStop}%
\bibitem [{\citenamefont {Scully} and \citenamefont
  {Zubairy}(1997)}]{scully1997quantum}%
  \BibitemOpen
  \bibfield  {author} {\bibinfo {author} {\bibfnamefont {M.}~\bibnamefont
  {Scully}} and \bibinfo {author} {\bibfnamefont {M.}~\bibnamefont {Zubairy}},
  }\href {https://books.google.com/books?id=20ISsQCKKmQC} {\emph {\bibinfo
  {title} {Quantum Optics}}}, Quantum Optics (\bibinfo  {publisher} {Cambridge
  University Press}, \bibinfo {year} {1997})\BibitemShut {NoStop}%
\bibitem [{\citenamefont {Lopata} and \citenamefont
  {Neuhauser}(2009)}]{lopota2009multiscale}%
  \BibitemOpen
  \bibfield  {author} {\bibinfo {author} {\bibfnamefont {K.}~\bibnamefont
  {Lopata}} and \bibinfo {author} {\bibfnamefont {D.}~\bibnamefont
  {Neuhauser}}, }\bibfield  {title} {\enquote {\bibinfo {title} {{Multiscale
  Maxwell–Schrödinger modeling: A split field finite-difference time-domain
  approach to molecular nanopolaritonics}},} }\href {\doibase
  10.1063/1.3082245} {\bibfield  {journal} {\bibinfo  {journal} {J. Chem.
  Phys.} }\textbf {\bibinfo {volume} {130}}, \bibinfo {pages} {104707}
  (\bibinfo {year} {2009})}\BibitemShut {NoStop}%
\bibitem [{\citenamefont {\ifmmode~\check{S}\else
  \v{S}\fi{}indelka}(2010)}]{sindelka2010derivation}%
  \BibitemOpen
  \bibfield  {author} {\bibinfo {author} {\bibfnamefont {M.}~\bibnamefont
  {\ifmmode~\check{S}\else \v{S}\fi{}indelka}}, }\bibfield  {title} {\enquote
  {\bibinfo {title} {Derivation of coupled {M}axwell-{S}chr\"odinger equations
  describing matter-laser interaction from first principles of quantum
  electrodynamics},} }\href {\doibase 10.1103/PhysRevA.81.033833} {\bibfield
  {journal} {\bibinfo  {journal} {Phys. Rev. A} }\textbf {\bibinfo {volume}
  {81}}, \bibinfo {pages} {033833} (\bibinfo {year} {2010})}\BibitemShut
  {NoStop}%
\bibitem [{\citenamefont {Sukharev} and \citenamefont
  {Nitzan}(2011)}]{sukharev2011numerical}%
  \BibitemOpen
  \bibfield  {author} {\bibinfo {author} {\bibfnamefont {M.}~\bibnamefont
  {Sukharev}} and \bibinfo {author} {\bibfnamefont {A.}~\bibnamefont {Nitzan}},
  }\bibfield  {title} {\enquote {\bibinfo {title} {Numerical studies of the
  interaction of an atomic sample with the electromagnetic field in two
  dimensions},} }\href {\doibase 10.1103/PhysRevA.84.043802} {\bibfield
  {journal} {\bibinfo  {journal} {Phys. Rev. A} }\textbf {\bibinfo {volume}
  {84}}, \bibinfo {pages} {043802} (\bibinfo {year} {2011})}\BibitemShut
  {NoStop}%
\bibitem [{\citenamefont {Li} \emph {et~al.}(2018)\citenamefont {Li},
  \citenamefont {Nitzan}, \citenamefont {Sukharev}, \citenamefont {Martinez},
  \citenamefont {Chen}, and \citenamefont {Subotnik}}]{li2018mixed}%
  \BibitemOpen
  \bibfield  {author} {\bibinfo {author} {\bibfnamefont {T.~E.} \bibnamefont
  {Li}}, \bibinfo {author} {\bibfnamefont {A.}~\bibnamefont {Nitzan}}, \bibinfo
  {author} {\bibfnamefont {M.}~\bibnamefont {Sukharev}}, \bibinfo {author}
  {\bibfnamefont {T.}~\bibnamefont {Martinez}}, \bibinfo {author}
  {\bibfnamefont {H.-T.} \bibnamefont {Chen}},  and \bibinfo {author}
  {\bibfnamefont {J.~E.} \bibnamefont {Subotnik}}, }\bibfield  {title}
  {\enquote {\bibinfo {title} {Mixed quantum-classical electrodynamics:
  Understanding spontaneous decay and zero-point energy},} }\href {\doibase
  10.1103/PhysRevA.97.032105} {\bibfield  {journal} {\bibinfo  {journal} {Phys.
  Rev. A} }\textbf {\bibinfo {volume} {97}}, \bibinfo {pages} {032105}
  (\bibinfo {year} {2018})}\BibitemShut {NoStop}%
\bibitem [{\citenamefont {Chen} \emph {et~al.}(2019)\citenamefont {Chen},
  \citenamefont {Li}, \citenamefont {Sukharev}, \citenamefont {Nitzan}, and
  \citenamefont {Subotnik}}]{chen2019ehrenfest}%
  \BibitemOpen
  \bibfield  {author} {\bibinfo {author} {\bibfnamefont {H.-T.} \bibnamefont
  {Chen}}, \bibinfo {author} {\bibfnamefont {T.~E.} \bibnamefont {Li}},
  \bibinfo {author} {\bibfnamefont {M.}~\bibnamefont {Sukharev}}, \bibinfo
  {author} {\bibfnamefont {A.}~\bibnamefont {Nitzan}},  and \bibinfo {author}
  {\bibfnamefont {J.~E.} \bibnamefont {Subotnik}}, }\bibfield  {title}
  {\enquote {\bibinfo {title} {{Ehrenfest+R dynamics. I. A mixed
  quantum–classical electrodynamics simulation of spontaneous emission}},}
  }\href {\doibase 10.1063/1.5057365} {\bibfield  {journal} {\bibinfo
  {journal} {J. Chem. Phys.} }\textbf {\bibinfo {volume} {150}}, \bibinfo
  {pages} {044102} (\bibinfo {year} {2019})}\BibitemShut {NoStop}%
\bibitem [{\citenamefont {Bhat} and \citenamefont
  {Sipe}(2001)}]{bhat2001optical}%
  \BibitemOpen
  \bibfield  {author} {\bibinfo {author} {\bibfnamefont {N.~A.~R.} \bibnamefont
  {Bhat}} and \bibinfo {author} {\bibfnamefont {J.~E.} \bibnamefont {Sipe}},
  }\bibfield  {title} {\enquote {\bibinfo {title} {Optical pulse propagation in
  nonlinear photonic crystals},} }\href {\doibase 10.1103/PhysRevE.64.056604}
  {\bibfield  {journal} {\bibinfo  {journal} {Phys. Rev. E} }\textbf {\bibinfo
  {volume} {64}}, \bibinfo {pages} {056604} (\bibinfo {year}
  {2001})}\BibitemShut {NoStop}%
\bibitem [{\citenamefont {Bhat} and \citenamefont
  {Sipe}(2006)}]{bhat2006hamiltonian}%
  \BibitemOpen
  \bibfield  {author} {\bibinfo {author} {\bibfnamefont {N.~A.~R.} \bibnamefont
  {Bhat}} and \bibinfo {author} {\bibfnamefont {J.~E.} \bibnamefont {Sipe}},
  }\bibfield  {title} {\enquote {\bibinfo {title} {Hamiltonian treatment of the
  electromagnetic field in dispersive and absorptive structured media},} }\href
  {\doibase 10.1103/PhysRevA.73.063808} {\bibfield  {journal} {\bibinfo
  {journal} {Phys. Rev. A} }\textbf {\bibinfo {volume} {73}}, \bibinfo {pages}
  {063808} (\bibinfo {year} {2006})}\BibitemShut {NoStop}%
\bibitem [{sm()}]{sm}%
  \BibitemOpen
  \href@noop {} {\emph {\bibinfo {title} {See Supplemental Material at [URL wil
  be inserted by publisher] for details on the mapping to Liouville space, the
  derivation of the pump-probe differential transmission, the phase expansion,
  as well as extension to multi-level systems. In addition, it includes
  Refs.~\cite{sommer2021molecular, schwennicke2024extracting,
  chen2022interplay}}}}\BibitemShut {NoStop}%
\bibitem [{Note1()}]{Note1}%
  \BibitemOpen
  \bibinfo {note} {Note that the $\rho ^{(n)}$ are not true density matrices,
  i.e., they do not obey the properties of density matrices such as the
  conservation of the trace}\BibitemShut {NoStop}%
\bibitem [{\citenamefont {Horn} and \citenamefont
  {Johnson}(1994)}]{horn1994topics}%
  \BibitemOpen
  \bibfield  {author} {\bibinfo {author} {\bibfnamefont {R.}~\bibnamefont
  {Horn}} and \bibinfo {author} {\bibfnamefont {C.}~\bibnamefont {Johnson}},
  }\href {https://books.google.com/books?id=LeuNXB2bl5EC} {\emph {\bibinfo
  {title} {Topics in Matrix Analysis}}} (\bibinfo  {publisher} {Cambridge
  University Press}, \bibinfo {year} {1994})\BibitemShut {NoStop}%
\bibitem [{\citenamefont {Am-Shallem} \emph {et~al.}(2015)\citenamefont
  {Am-Shallem}, \citenamefont {Levy}, \citenamefont {Schaefer}, and
  \citenamefont {Kosloff}}]{amshallem2015approaches}%
  \BibitemOpen
  \bibfield  {author} {\bibinfo {author} {\bibfnamefont {M.}~\bibnamefont
  {Am-Shallem}}, \bibinfo {author} {\bibfnamefont {A.}~\bibnamefont {Levy}},
  \bibinfo {author} {\bibfnamefont {I.}~\bibnamefont {Schaefer}},  and \bibinfo
  {author} {\bibfnamefont {R.}~\bibnamefont {Kosloff}}, }\href@noop {}
  {\enquote {\bibinfo {title} {Three approaches for representing {L}indblad
  dynamics by a matrix-vector notation},} } (\bibinfo {year} {2015}), \Eprint
  {http://arxiv.org/abs/1510.08634} {arXiv:1510.08634 [quant-ph]} \BibitemShut
  {NoStop}%
\bibitem [{\citenamefont {Rocca} \emph {et~al.}(1998)\citenamefont {Rocca},
  \citenamefont {Bassani}, and \citenamefont
  {Agranovich}}]{larocca1998biexcitons}%
  \BibitemOpen
  \bibfield  {author} {\bibinfo {author} {\bibfnamefont {G.~C.~L.} \bibnamefont
  {Rocca}}, \bibinfo {author} {\bibfnamefont {F.}~\bibnamefont {Bassani}},  and
  \bibinfo {author} {\bibfnamefont {V.~M.} \bibnamefont {Agranovich}},
  }\bibfield  {title} {\enquote {\bibinfo {title} {Biexcitons and dark states
  in semiconductor microcavities},} }\href {\doibase 10.1364/JOSAB.15.000652}
  {\bibfield  {journal} {\bibinfo  {journal} {J. Opt. Soc. Am. B} }\textbf
  {\bibinfo {volume} {15}}, \bibinfo {pages} {652--660} (\bibinfo {year}
  {1998})}\BibitemShut {NoStop}%
\bibitem [{\citenamefont {Agranovich} \emph {et~al.}(2003)\citenamefont
  {Agranovich}, \citenamefont {Litinskaia}, and \citenamefont
  {Lidzey}}]{agranovic2003cavity}%
  \BibitemOpen
  \bibfield  {author} {\bibinfo {author} {\bibfnamefont {V.~M.} \bibnamefont
  {Agranovich}}, \bibinfo {author} {\bibfnamefont {M.}~\bibnamefont
  {Litinskaia}},  and \bibinfo {author} {\bibfnamefont {D.~G.} \bibnamefont
  {Lidzey}}, }\bibfield  {title} {\enquote {\bibinfo {title} {Cavity polaritons
  in microcavities containing disordered organic semiconductors},} }\href
  {\doibase 10.1103/PhysRevB.67.085311} {\bibfield  {journal} {\bibinfo
  {journal} {Phys. Rev. B} }\textbf {\bibinfo {volume} {67}}, \bibinfo {pages}
  {085311} (\bibinfo {year} {2003})}\BibitemShut {NoStop}%
\bibitem [{\citenamefont {Litinskaya} \emph {et~al.}(2004)\citenamefont
  {Litinskaya}, \citenamefont {Reineker}, and \citenamefont
  {Agranovich}}]{litinskaya2004fast}%
  \BibitemOpen
  \bibfield  {author} {\bibinfo {author} {\bibfnamefont {M.}~\bibnamefont
  {Litinskaya}}, \bibinfo {author} {\bibfnamefont {P.}~\bibnamefont
  {Reineker}},  and \bibinfo {author} {\bibfnamefont {V.}~\bibnamefont
  {Agranovich}}, }\bibfield  {title} {\enquote {\bibinfo {title} {Fast
  polariton relaxation in strongly coupled organic microcavities},} }\href
  {\doibase https://doi.org/10.1016/j.jlumin.2004.08.033} {\bibfield  {journal}
  {\bibinfo  {journal} {J. Lumin.} }\textbf {\bibinfo {volume} {110}}, \bibinfo
  {pages} {364--372} (\bibinfo {year} {2004})}\BibitemShut {NoStop}%
\bibitem [{\citenamefont {Gonzalez-Ballestero} \emph
  {et~al.}(2016)\citenamefont {Gonzalez-Ballestero}, \citenamefont {Feist},
  \citenamefont {Gonzalo~Bad\'{\i}a}, \citenamefont {Moreno}, and \citenamefont
  {Garcia-Vidal}}]{gonzalezballestero2016uncoupled}%
  \BibitemOpen
  \bibfield  {author} {\bibinfo {author} {\bibfnamefont {C.}~\bibnamefont
  {Gonzalez-Ballestero}}, \bibinfo {author} {\bibfnamefont {J.}~\bibnamefont
  {Feist}}, \bibinfo {author} {\bibfnamefont {E.}~\bibnamefont
  {Gonzalo~Bad\'{\i}a}}, \bibinfo {author} {\bibfnamefont {E.}~\bibnamefont
  {Moreno}},  and \bibinfo {author} {\bibfnamefont {F.~J.} \bibnamefont
  {Garcia-Vidal}}, }\bibfield  {title} {\enquote {\bibinfo {title} {Uncoupled
  dark states can inherit polaritonic properties},} }\href {\doibase
  10.1103/PhysRevLett.117.156402} {\bibfield  {journal} {\bibinfo  {journal}
  {Phys. Rev. Lett.} }\textbf {\bibinfo {volume} {117}}, \bibinfo {pages}
  {156402} (\bibinfo {year} {2016})}\BibitemShut {NoStop}%
\bibitem [{\citenamefont {Campos-Gonzalez-Angulo} and \citenamefont
  {Yuen-Zhou}(2022)}]{camposgonzalez2022generalization}%
  \BibitemOpen
  \bibfield  {author} {\bibinfo {author} {\bibfnamefont {J.~A.} \bibnamefont
  {Campos-Gonzalez-Angulo}} and \bibinfo {author} {\bibfnamefont
  {J.}~\bibnamefont {Yuen-Zhou}}, }\bibfield  {title} {\enquote {\bibinfo
  {title} {{Generalization of the Tavis–Cummings model for multi-level
  anharmonic systems: Insights on the second excitation manifold}},} }\href
  {\doibase 10.1063/5.0087234} {\bibfield  {journal} {\bibinfo  {journal} {J.
  Chem. Phys.} }\textbf {\bibinfo {volume} {156}}, \bibinfo {pages} {194308}
  (\bibinfo {year} {2022})}\BibitemShut {NoStop}%
\bibitem [{\citenamefont {Yuen-Zhou} and \citenamefont
  {Koner}(2024)}]{koner2024linear}%
  \BibitemOpen
  \bibfield  {author} {\bibinfo {author} {\bibfnamefont {J.}~\bibnamefont
  {Yuen-Zhou}} and \bibinfo {author} {\bibfnamefont {A.}~\bibnamefont {Koner}},
  }\bibfield  {title} {\enquote {\bibinfo {title} {{Linear response of
  molecular polaritons}},} }\href {\doibase 10.1063/5.0183683} {\bibfield
  {journal} {\bibinfo  {journal} {J. Chem. Phys.} }\textbf {\bibinfo {volume}
  {160}}, \bibinfo {pages} {154107} (\bibinfo {year} {2024})}\BibitemShut
  {NoStop}%
\bibitem [{\citenamefont {Steck}(2007)}]{steck2007quantum}%
  \BibitemOpen
  \bibfield  {author} {\bibinfo {author} {\bibfnamefont {D.}~\bibnamefont
  {Steck}}, }\href {https://books.google.com/books?id=bc9TMwEACAAJ} {\emph
  {\bibinfo {title} {Quantum and Atom Optics}}} (\bibinfo {year}
  {2007})\BibitemShut {NoStop}%
\bibitem [{\citenamefont {Ciuti} and \citenamefont
  {Carusotto}(2006)}]{ciuti2006input}%
  \BibitemOpen
  \bibfield  {author} {\bibinfo {author} {\bibfnamefont {C.}~\bibnamefont
  {Ciuti}} and \bibinfo {author} {\bibfnamefont {I.}~\bibnamefont {Carusotto}},
  }\bibfield  {title} {\enquote {\bibinfo {title} {Input-output theory of
  cavities in the ultrastrong coupling regime: The case of time-independent
  cavity parameters},} }\href {\doibase 10.1103/PhysRevA.74.033811} {\bibfield
  {journal} {\bibinfo  {journal} {Phys. Rev. A} }\textbf {\bibinfo {volume}
  {74}}, \bibinfo {pages} {033811} (\bibinfo {year} {2006})}\BibitemShut
  {NoStop}%
\bibitem [{\citenamefont {Frisk~Kockum} \emph {et~al.}(2019)\citenamefont
  {Frisk~Kockum}, \citenamefont {Miranowicz}, \citenamefont {De~Liberato},
  \citenamefont {Savasta}, and \citenamefont
  {Nori}}]{friskkockum2019ultrastrong}%
  \BibitemOpen
  \bibfield  {author} {\bibinfo {author} {\bibfnamefont {A.}~\bibnamefont
  {Frisk~Kockum}}, \bibinfo {author} {\bibfnamefont {A.}~\bibnamefont
  {Miranowicz}}, \bibinfo {author} {\bibfnamefont {S.}~\bibnamefont
  {De~Liberato}}, \bibinfo {author} {\bibfnamefont {S.}~\bibnamefont
  {Savasta}},  and \bibinfo {author} {\bibfnamefont {F.}~\bibnamefont {Nori}},
  }\bibfield  {title} {\enquote {\bibinfo {title} {Ultrastrong coupling between
  light and matter},} }\href {\doibase 10.1038/s42254-018-0006-2} {\bibfield
  {journal} {\bibinfo  {journal} {Nat. Rev. Phys.} }\textbf {\bibinfo {volume}
  {1}}, \bibinfo {pages} {19--40} (\bibinfo {year} {2019})}\BibitemShut
  {NoStop}%
\bibitem [{\citenamefont {Hybl} \emph {et~al.}(1998)\citenamefont {Hybl},
  \citenamefont {Albrecht}, \citenamefont {{Gallagher Faeder}}, and
  \citenamefont {Jonas}}]{hybl1998two}%
  \BibitemOpen
  \bibfield  {author} {\bibinfo {author} {\bibfnamefont {J.~D.} \bibnamefont
  {Hybl}}, \bibinfo {author} {\bibfnamefont {A.~W.} \bibnamefont {Albrecht}},
  \bibinfo {author} {\bibfnamefont {S.~M.} \bibnamefont {{Gallagher Faeder}}},
  and \bibinfo {author} {\bibfnamefont {D.~M.} \bibnamefont {Jonas}},
  }\bibfield  {title} {\enquote {\bibinfo {title} {Two-dimensional electronic
  spectroscopy},} }\href {\doibase
  https://doi.org/10.1016/S0009-2614(98)01140-3} {\bibfield  {journal}
  {\bibinfo  {journal} {Chem. Phys. Lett.} }\textbf {\bibinfo {volume} {297}},
  \bibinfo {pages} {307--313} (\bibinfo {year} {1998})}\BibitemShut {NoStop}%
\bibitem [{\citenamefont {Tian} \emph {et~al.}(2003)\citenamefont {Tian},
  \citenamefont {Keusters}, \citenamefont {Suzaki}, and \citenamefont
  {Warren}}]{tian2003femtosecond}%
  \BibitemOpen
  \bibfield  {author} {\bibinfo {author} {\bibfnamefont {P.}~\bibnamefont
  {Tian}}, \bibinfo {author} {\bibfnamefont {D.}~\bibnamefont {Keusters}},
  \bibinfo {author} {\bibfnamefont {Y.}~\bibnamefont {Suzaki}},  and \bibinfo
  {author} {\bibfnamefont {W.~S.} \bibnamefont {Warren}}, }\bibfield  {title}
  {\enquote {\bibinfo {title} {Femtosecond phase-coherent two-dimensional
  spectroscopy},} }\href {\doibase 10.1126/science.1083433} {\bibfield
  {journal} {\bibinfo  {journal} {Science} }\textbf {\bibinfo {volume} {300}},
  \bibinfo {pages} {1553--1555} (\bibinfo {year} {2003})}\BibitemShut {NoStop}%
\bibitem [{\citenamefont {Tan}(2008)}]{tan2008theory}%
  \BibitemOpen
  \bibfield  {author} {\bibinfo {author} {\bibfnamefont {H.-S.} \bibnamefont
  {Tan}}, }\bibfield  {title} {\enquote {\bibinfo {title} {{Theory and
  phase-cycling scheme selection principles of collinear phase coherent
  multi-dimensional optical spectroscopy}},} }\href {\doibase
  10.1063/1.2978381} {\bibfield  {journal} {\bibinfo  {journal} {J. Chem.
  Phys.} }\textbf {\bibinfo {volume} {129}}, \bibinfo {pages} {124501}
  (\bibinfo {year} {2008})}\BibitemShut {NoStop}%
\bibitem [{\citenamefont {Yang} \emph {et~al.}(2020)\citenamefont {Yang},
  \citenamefont {Xiang}, and \citenamefont {Xiong}}]{yang2020controlling}%
  \BibitemOpen
  \bibfield  {author} {\bibinfo {author} {\bibfnamefont {Z.}~\bibnamefont
  {Yang}}, \bibinfo {author} {\bibfnamefont {B.}~\bibnamefont {Xiang}},  and
  \bibinfo {author} {\bibfnamefont {W.}~\bibnamefont {Xiong}}, }\bibfield
  {title} {\enquote {\bibinfo {title} {Controlling quantum pathways in
  molecular vibrational polaritons},} }\href {\doibase
  10.1021/acsphotonics.0c00148} {\bibfield  {journal} {\bibinfo  {journal} {ACS
  Photonics} }\textbf {\bibinfo {volume} {7}}, \bibinfo {pages} {919--924}
  (\bibinfo {year} {2020})}\BibitemShut {NoStop}%
\bibitem [{\citenamefont {Fowler-Wright} \emph {et~al.}(2023)\citenamefont
  {Fowler-Wright}, \citenamefont {Arnard\'ottir}, \citenamefont {Kirton},
  \citenamefont {Lovett}, and \citenamefont
  {Keeling}}]{fowlerwright2023determining}%
  \BibitemOpen
  \bibfield  {author} {\bibinfo {author} {\bibfnamefont {P.}~\bibnamefont
  {Fowler-Wright}}, \bibinfo {author} {\bibfnamefont {K.~B.} \bibnamefont
  {Arnard\'ottir}}, \bibinfo {author} {\bibfnamefont {P.}~\bibnamefont
  {Kirton}}, \bibinfo {author} {\bibfnamefont {B.~W.} \bibnamefont {Lovett}},
  and \bibinfo {author} {\bibfnamefont {J.}~\bibnamefont {Keeling}}, }\bibfield
   {title} {\enquote {\bibinfo {title} {Determining the validity of cumulant
  expansions for central spin models},} }\href {\doibase
  10.1103/PhysRevResearch.5.033148} {\bibfield  {journal} {\bibinfo  {journal}
  {Phys. Rev. Res.} }\textbf {\bibinfo {volume} {5}}, \bibinfo {pages} {033148}
  (\bibinfo {year} {2023})}\BibitemShut {NoStop}%
\bibitem [{\citenamefont {Axt} and \citenamefont
  {Stahl}(1994)}]{axt1994dynamics}%
  \BibitemOpen
  \bibfield  {author} {\bibinfo {author} {\bibfnamefont {V.~M.} \bibnamefont
  {Axt}} and \bibinfo {author} {\bibfnamefont {A.}~\bibnamefont {Stahl}},
  }\bibfield  {title} {\enquote {\bibinfo {title} {A dynamics-controlled
  truncation scheme for the hierarchy of density matrices in semiconductor
  optics},} }\href {\doibase 10.1007/BF01316963} {\bibfield  {journal}
  {\bibinfo  {journal} {Z. Phys. B Cond. Matt.} }\textbf {\bibinfo {volume}
  {93}}, \bibinfo {pages} {195--204} (\bibinfo {year} {1994})}\BibitemShut
  {NoStop}%
\bibitem [{\citenamefont {Gruetzmacher} and \citenamefont
  {Scherer}(2003)}]{gruetzmacher2003finite}%
  \BibitemOpen
  \bibfield  {author} {\bibinfo {author} {\bibfnamefont {J.~A.} \bibnamefont
  {Gruetzmacher}} and \bibinfo {author} {\bibfnamefont {N.~F.} \bibnamefont
  {Scherer}}, }\bibfield  {title} {\enquote {\bibinfo {title}
  {Finite-difference time-domain simulation of ultrashort pulse propagation
  incorporating quantum-mechanical response functions},} }\href {\doibase
  10.1364/OL.28.000573} {\bibfield  {journal} {\bibinfo  {journal} {Opt. Lett.}
  }\textbf {\bibinfo {volume} {28}}, \bibinfo {pages} {573--575} (\bibinfo
  {year} {2003})}\BibitemShut {NoStop}%
\bibitem [{\citenamefont {Zhou} \emph {et~al.}(2024)\citenamefont {Zhou},
  \citenamefont {Gangaraj}, \citenamefont {Zhou}, and \citenamefont
  {Yu}}]{zhou2024simulating}%
  \BibitemOpen
  \bibfield  {author} {\bibinfo {author} {\bibfnamefont {Q.}~\bibnamefont
  {Zhou}}, \bibinfo {author} {\bibfnamefont {S.~A.~H.} \bibnamefont
  {Gangaraj}}, \bibinfo {author} {\bibfnamefont {M.}~\bibnamefont {Zhou}},  and
  \bibinfo {author} {\bibfnamefont {Z.}~\bibnamefont {Yu}}, }\href@noop {}
  {\enquote {\bibinfo {title} {Simulating quantum emitters in arbitrary
  photonic environments using {F}{D}{T}{D}: beyond the semi-classical regime},}
  } (\bibinfo {year} {2024}), \Eprint {http://arxiv.org/abs/2410.16118}
  {arXiv:2410.16118 [quant-ph]} \BibitemShut {NoStop}%
\bibitem [{\citenamefont {Yang} \emph {et~al.}(2023)\citenamefont {Yang},
  \citenamefont {Bhakta}, and \citenamefont {Xiong}}]{yang2023enabling}%
  \BibitemOpen
  \bibfield  {author} {\bibinfo {author} {\bibfnamefont {Z.}~\bibnamefont
  {Yang}}, \bibinfo {author} {\bibfnamefont {H.~H.} \bibnamefont {Bhakta}},
  and \bibinfo {author} {\bibfnamefont {W.}~\bibnamefont {Xiong}}, }\bibfield
  {title} {\enquote {\bibinfo {title} {Enabling multiple intercavity polariton
  coherences by adding quantum confinement to cavity molecular polaritons},}
  }\href {\doibase 10.1073/pnas.2206062120} {\bibfield  {journal} {\bibinfo
  {journal} {Proc. Natl. Acad. Sci. USA} }\textbf {\bibinfo {volume} {120}},
  \bibinfo {pages} {e2206062120} (\bibinfo {year} {2023})}\BibitemShut
  {NoStop}%
\bibitem [{\citenamefont {Balasubrahmaniyam} \emph {et~al.}(2023)\citenamefont
  {Balasubrahmaniyam}, \citenamefont {Simkhovich}, \citenamefont {Golombek},
  \citenamefont {Sandik}, \citenamefont {Ankonina}, and \citenamefont
  {Schwartz}}]{balasubrahmaniyam2023from}%
  \BibitemOpen
  \bibfield  {author} {\bibinfo {author} {\bibfnamefont {M.}~\bibnamefont
  {Balasubrahmaniyam}}, \bibinfo {author} {\bibfnamefont {A.}~\bibnamefont
  {Simkhovich}}, \bibinfo {author} {\bibfnamefont {A.}~\bibnamefont
  {Golombek}}, \bibinfo {author} {\bibfnamefont {G.}~\bibnamefont {Sandik}},
  \bibinfo {author} {\bibfnamefont {G.}~\bibnamefont {Ankonina}},  and \bibinfo
  {author} {\bibfnamefont {T.}~\bibnamefont {Schwartz}}, }\bibfield  {title}
  {\enquote {\bibinfo {title} {From enhanced diffusion to ultrafast ballistic
  motion of hybrid light--matter excitations},} }\href {\doibase
  10.1038/s41563-022-01463-3} {\bibfield  {journal} {\bibinfo  {journal} {Nat.
  Mater.} }\textbf {\bibinfo {volume} {22}}, \bibinfo {pages} {338--344}
  (\bibinfo {year} {2023})}\BibitemShut {NoStop}%
\bibitem [{\citenamefont {Xu} \emph {et~al.}(2023)\citenamefont {Xu},
  \citenamefont {Mandal}, \citenamefont {Baxter}, \citenamefont {Cheng},
  \citenamefont {Lee}, \citenamefont {Su}, \citenamefont {Liu}, \citenamefont
  {Reichman}, and \citenamefont {Delor}}]{xu2023ultrafast}%
  \BibitemOpen
  \bibfield  {author} {\bibinfo {author} {\bibfnamefont {D.}~\bibnamefont
  {Xu}}, \bibinfo {author} {\bibfnamefont {A.}~\bibnamefont {Mandal}}, \bibinfo
  {author} {\bibfnamefont {J.~M.} \bibnamefont {Baxter}}, \bibinfo {author}
  {\bibfnamefont {S.-W.} \bibnamefont {Cheng}}, \bibinfo {author}
  {\bibfnamefont {I.}~\bibnamefont {Lee}}, \bibinfo {author} {\bibfnamefont
  {H.}~\bibnamefont {Su}}, \bibinfo {author} {\bibfnamefont {S.}~\bibnamefont
  {Liu}}, \bibinfo {author} {\bibfnamefont {D.~R.} \bibnamefont {Reichman}},
  and \bibinfo {author} {\bibfnamefont {M.}~\bibnamefont {Delor}}, }\bibfield
  {title} {\enquote {\bibinfo {title} {Ultrafast imaging of polariton
  propagation and interactions},} }\href {\doibase 10.1038/s41467-023-39550-x}
  {\bibfield  {journal} {\bibinfo  {journal} {Nat. Commun.} }\textbf {\bibinfo
  {volume} {14}}, \bibinfo {pages} {3881} (\bibinfo {year} {2023})}\BibitemShut
  {NoStop}%
\bibitem [{\citenamefont {Finkelstein-Shapiro} \emph
  {et~al.}(2023)\citenamefont {Finkelstein-Shapiro}, \citenamefont {Mante},
  \citenamefont {Balci}, \citenamefont {Zigmantas}, and \citenamefont
  {Pullerits}}]{finkelstein2023non}%
  \BibitemOpen
  \bibfield  {author} {\bibinfo {author} {\bibfnamefont {D.}~\bibnamefont
  {Finkelstein-Shapiro}}, \bibinfo {author} {\bibfnamefont {P.-A.} \bibnamefont
  {Mante}}, \bibinfo {author} {\bibfnamefont {S.}~\bibnamefont {Balci}},
  \bibinfo {author} {\bibfnamefont {D.}~\bibnamefont {Zigmantas}},  and
  \bibinfo {author} {\bibfnamefont {T.}~\bibnamefont {Pullerits}}, }\bibfield
  {title} {\enquote {\bibinfo {title} {{Non-Hermitian Hamiltonians for linear
  and nonlinear optical response: A model for plexcitons}},} }\href {\doibase
  10.1063/5.0130287} {\bibfield  {journal} {\bibinfo  {journal} {J. Chem.
  Phys.} }\textbf {\bibinfo {volume} {158}}, \bibinfo {pages} {104104}
  (\bibinfo {year} {2023})}\BibitemShut {NoStop}%
\bibitem [{\citenamefont {Fryett} \emph {et~al.}(2018)\citenamefont {Fryett},
  \citenamefont {Zhan}, and \citenamefont {Majumdar}}]{fryett2018cavity}%
  \BibitemOpen
  \bibfield  {author} {\bibinfo {author} {\bibfnamefont {T.}~\bibnamefont
  {Fryett}}, \bibinfo {author} {\bibfnamefont {A.}~\bibnamefont {Zhan}},  and
  \bibinfo {author} {\bibfnamefont {A.}~\bibnamefont {Majumdar}}, }\bibfield
  {title} {\enquote {\bibinfo {title} {Cavity nonlinear optics with layered
  materials},} }\href {\doibase doi:10.1515/nanoph-2017-0069} {\bibfield
  {journal} {\bibinfo  {journal} {Nanophotonics} }\textbf {\bibinfo {volume}
  {7}}, \bibinfo {pages} {355--370} (\bibinfo {year} {2018})}\BibitemShut
  {NoStop}%
\bibitem [{\citenamefont {Sommer} \emph {et~al.}(2021)\citenamefont {Sommer},
  \citenamefont {Reitz}, \citenamefont {Mineo}, and \citenamefont
  {Genes}}]{sommer2021molecular}%
  \BibitemOpen
  \bibfield  {author} {\bibinfo {author} {\bibfnamefont {C.}~\bibnamefont
  {Sommer}}, \bibinfo {author} {\bibfnamefont {M.}~\bibnamefont {Reitz}},
  \bibinfo {author} {\bibfnamefont {F.}~\bibnamefont {Mineo}},  and \bibinfo
  {author} {\bibfnamefont {C.}~\bibnamefont {Genes}}, }\bibfield  {title}
  {\enquote {\bibinfo {title} {Molecular polaritonics in dense mesoscopic
  disordered ensembles},} }\href {\doibase 10.1103/PhysRevResearch.3.033141}
  {\bibfield  {journal} {\bibinfo  {journal} {Phys. Rev. Res.} }\textbf
  {\bibinfo {volume} {3}}, \bibinfo {pages} {033141} (\bibinfo {year}
  {2021})}\BibitemShut {NoStop}%
\bibitem [{\citenamefont {Schwennicke} \emph {et~al.}(2024)\citenamefont
  {Schwennicke}, \citenamefont {Giebink}, and \citenamefont
  {Yuen-Zhou}}]{schwennicke2024extracting}%
  \BibitemOpen
  \bibfield  {author} {\bibinfo {author} {\bibfnamefont {K.}~\bibnamefont
  {Schwennicke}}, \bibinfo {author} {\bibfnamefont {N.~C.} \bibnamefont
  {Giebink}},  and \bibinfo {author} {\bibfnamefont {J.}~\bibnamefont
  {Yuen-Zhou}}, }\bibfield  {title} {\enquote {\bibinfo {title} {Extracting
  accurate light–matter couplings from disordered polaritons},} }\href
  {\doibase doi:10.1515/nanoph-2024-0049} {\bibfield  {journal} {\bibinfo
  {journal} {Nanophotonics} }\textbf {\bibinfo {volume} {13}}, \bibinfo {pages}
  {2469--2478} (\bibinfo {year} {2024})}\BibitemShut {NoStop}%
\bibitem [{\citenamefont {Chen} \emph {et~al.}(2022)\citenamefont {Chen},
  \citenamefont {Zhou}, \citenamefont {Sukharev}, \citenamefont {Subotnik}, and
  \citenamefont {Nitzan}}]{chen2022interplay}%
  \BibitemOpen
  \bibfield  {author} {\bibinfo {author} {\bibfnamefont {H.-T.} \bibnamefont
  {Chen}}, \bibinfo {author} {\bibfnamefont {Z.}~\bibnamefont {Zhou}}, \bibinfo
  {author} {\bibfnamefont {M.}~\bibnamefont {Sukharev}}, \bibinfo {author}
  {\bibfnamefont {J.~E.} \bibnamefont {Subotnik}},  and \bibinfo {author}
  {\bibfnamefont {A.}~\bibnamefont {Nitzan}}, }\bibfield  {title} {\enquote
  {\bibinfo {title} {Interplay between disorder and collective coherent
  response: Superradiance and spectral motional narrowing in the time domain},}
  }\href {\doibase 10.1103/PhysRevA.106.053703} {\bibfield  {journal} {\bibinfo
   {journal} {Phys. Rev. A} }\textbf {\bibinfo {volume} {106}}, \bibinfo
  {pages} {053703} (\bibinfo {year} {2022})}\BibitemShut {NoStop}%
\end{thebibliography}%
\newpage  % Ensure a page break before switching to one-column layout
\onecolumngrid  % Start one-column grid
\clearpage

% Add a symmetrically centered title to the appendix
\begin{center}
    \large\textbf{Supplemental Material to \textit{Nonlinear semiclassical spectroscopy of ultrafast molecular polariton dynamics}}
\end{center}

% Redefine section numbering to include 'S'
\renewcommand{\thesection}{S\arabic{section}}

% Redefine equation numbering to include 'S'
\renewcommand{\theequation}{S\arabic{equation}}

% Reset the equation counter for the appendix
\setcounter{equation}{0}

% Redefine figure numbering to include 'S'
\renewcommand{\thefigure}{S\arabic{figure}}

% Reset the figure counter for the appendix
\setcounter{figure}{0}

\section{Vectorization/mapping to Liouville space}

It is advantageous to convert the von-Neumann (master) equation for the density matrix onto matrix-vector form, i.e., map the $n\times n$ density matrix $\rho$ onto an $n^2$ vector  $\vec\rho $ and the superoperator acting on the density matrix onto an $n^2\times n^2$ matrix $\mathcal L$ (referred to as Liouville space). For instance, for a two-level system, the density matrix is mapped onto a column vector as
\begin{align}
\rho=\begin{pmatrix}
\rho_{11} & \rho_{12} \\
\rho_{21} & \rho_{22}
\end{pmatrix}
\quad \to \quad 
\vec{\rho}\equiv|\rho\rangle\rangle=\begin{pmatrix}
\rho_{11}\\
\rho_{12}\\
\rho_{21}\\
\rho_{22}
\end{pmatrix}.
\end{align}
This can be done in a more formal way in a procedure known in the literature as vec-ing \cite{amshallem2015approaches, horn1994topics}, which has the following rules
\begin{enumerate}
\item A left multiplication of the matrix $\rho$ by an $n\times n$ matrix $A$, i.e., $A\rho$, is equivalent
to an operation on the vector $\vec\rho$ by the $n^2\times n^2$ matrix $A\otimes I$, where $I$ is the $n \times n$
identity matrix, and $\otimes$ is the Kronecker product.
\item Similarly, a right multiplication of the matrix $\rho$ by an $n\times n$ matrix $B$, i.e., $\rho B$, is
equivalent to an operation on the vector $\vec\rho$ by the $n^2\times n^2$ matrix $I \otimes B^\top$. 
\item Finally, a combination of left and right matrices multiplication, $A\rho B$, is equivalent
to an operation on the vector $\vec\rho$ by the $n^2\times n^2$ matrix $A \otimes B^\top$.
\end{enumerate}
With this, one can express general commutators as
\begin{align}
[\mathcal H, \rho]\to\left(\mathcal H\otimes I  - I\otimes \mathcal H^\top\right)\vec\rho \equiv \mathcal L \vec\rho.
\end{align}

\section{Cavity input-output relations}
\label{sec:inputoutput}

Let us consider a (classical) input field $\expval{\hat b_{\mathrm{in}, L}(t)}$ driving the cavity through the left mirror described by the Hamiltonian
\begin{align}
\mathcal{H}_d=-\mi \hbar \sqrt{\kappa_L}\expval{\hat b_{\mathrm{in}, L}(t)} \hat a^\dagger+\mathrm{H.c.},
\end{align}
where $\kappa_L$ describes the loss rate through the left mirror. For our purposes, we consider an input pulse described by $\expval{\hat b_{\mathrm{in}, L}(t)}=\eta f(t)/\sqrt{\kappa_L}$ (in a frame rotating at the pulse frequency). The parameter $\eta$ can be related to the incident laser power $\mathcal P\ts{in}$ as $\eta=\sqrt{\mathcal{P}\ts{in}\kappa_L/(\hbar\omega_\ell)}$, where $\omega_\ell$ is the pulse carrier frequency.
Then, the cavity input-output relations written separately at the left ($L$) and right ($R$) mirror, respectively, are given by \cite{steck2007quantum}
\begin{subequations}
\begin{align}
\hat b_{\mathrm{out}, L}(t)-\hat b_{\mathrm{in}, L}(t)&=\sqrt{\kappa_L}\hat{a},\\
\hat b_{\mathrm{out}, R}(t)-\hat b_{\mathrm{in}, R}(t)&=\sqrt{\kappa_R}\hat{a},
\end{align}
\end{subequations}
from which the transmitted intensity can be straightforwardly derived as 
\begin{align}
T=\left|\frac{\expval{\hat b_{\mathrm{out}, R}(t)}}{\expval{\hat b_{\mathrm{in}, L}(t)}} \right| ^2=\frac{\kappa_R\kappa_L}{\eta^2 f(t)^2}|\alpha |^2.
\end{align}
For a balanced cavity $\kappa_L=\kappa_R=\kappa/2$, this becomes
\begin{align}
T=\frac{(\kappa/2)^2}{\eta^2 f(t)^2}|\alpha |^2,
\end{align}
which holds in both linear and nonlinear regimes, as well as in the Fourier domain. \revise{It is important to note, however, that in the ultrastrong coupling regime, the input-output relations are significantly modified due to the breakdown of the rotating wave approximation (RWA). In this regime, where the collective coupling strength $g\sqrt{\mathcal{N}}$ becomes comparable to the cavity or matter transition frequencies, non-trivial corrections to the input-output relations must be incorporated that account for counter-rotating terms and the hybrid eigenmodes of the light-matter system \cite{ciuti2006input, friskkockum2019ultrastrong}.}

\section{General form of pump-probe equations}

For a cavity driven by two pulses, the $(n)(m)$-th order equations for the field and molecular density matrix are given by (assuming initial conditions $\alpha(0)=\alpha^{(0)(0)}=0$)
\begin{subequations}
\begin{align}
\dot{\alpha}^{(n) (m)} &=-\left(\frac{\kappa}{2}+\mi\omega_c\right)\alpha^{(n)(m)}-\mi E_0\mathcal{N}\mathrm{Tr}[\hat{\mu}\rho^{(n)(m)}]-\delta_{1n}\delta_{0m} f_p (t-\tau_p)\me^{-\mi\omega_p t}-\delta_{0n}\delta_{1m}f_{p'}(t-\tau_{p'})\me^{-\mi\omega_{p'} t},\\
\dot{\rho}^{(n) (m)}(t) &=-\mi[\mathcal{H}_0,\rho^{(n) (m)}]-\mi E_0 \sum_{j=0}^{n-1}\sum_{j'=0}^{m-1}\left(\alpha^{(n-j)(m-j')}+\alpha^{(n-j)(m-j')*}\right)\left[\hat{\mu}, \rho^{(j) (j')}\right],
\end{align}
\end{subequations}
where we will generally assume identical pulse envelopes $f_j(t-\tau_j)=\me^{-(t-\tau_j)^2/\tau_w^2}$. In Liouville space, the equation for the density vector can then be expressed in compact form as
\begin{align}
\label{eq:pumpprobedensity}
\dot{\vec{\rho}}^{(n)(m)} (t)=-\mi\mathcal{L}_0\vec{\rho}^{(n)(m)} (t)-\mi \sum_{j=0}^{n-1}\sum_{j'=0}^{m-1}\mathcal{L}\ts{int}^{(n-j)(m-j')}(t)\vec{\rho}^{(j)(j')}(t),
\end{align}
where we defined the interaction Liouvillian
\begin{align}
\mathcal{L}\ts{int}^{(n-j)(m-j')}(t)=E_0[\alpha^{(n-j)(m-j')}(t)+\mathrm{c.c.}]\mathcal{L}_\mu,
\end{align}
and the vectorized commutators are given by
\begin{subequations}
\begin{align}
[\mathcal{H}_0,\rho]\to\mathcal{L}_0\vec{\rho} &= [\mathcal{H}_0\otimes I-I\otimes \mathcal{H}_0^\top]\vec{\rho} ,\\
\label{eq:dipolecommutator}
[\hat\mu,\rho]\to\mathcal{L}_\mu\vec{\rho}&= [\hat{\mu}\otimes I-I\otimes \hat{\mu}^\top ]\vec{\rho}.
\end{align}
\end{subequations}

\section{Differential transmission}
\label{sec:difftransmission}

We define the differential transmission (DT) as the difference in the probe transmission with and without the presence of the pump pulse
\begin{align}
\Delta T (\omega)=T_{p'}^\text{pump\, on}(\omega)-T_{p'}^\text{pump\, off}(\omega).
\end{align}
 Considering only the lowest (third) order correction of the probe field due to the pump, the normalized transmitted probe intensities can be expressed from input-output relations in Fourier domain (see section \ref{sec:inputoutput}) as 
\begin{align}
T_{p'}^\text{pump\, off}(\omega)=\left(\frac{\kappa}{2}\right)^2\frac{|\alpha^{(0)(1)}(\omega)|^2}{f_{p'}(\omega)^2},\qquad  T_{p'}^\text{pump\, on}(\omega)\approx\left(\frac{\kappa}{2}\right)^2\frac{|\alpha^{(0)(1)}(\omega)+\eta_p
^2\alpha^{(2)(1)}(\omega)|^2}{f_{p'}(\omega)^2}.
\end{align}
From this, we can approximate the DT as
\begin{align}
\Delta T (\omega)\approx\left(\frac{\kappa}{2}\right)^2\eta_p^2\frac{2\mathrm{Re}\left[\alpha^{(0)(1)*}(\omega)\alpha^{(2)(1)}(\omega)\right]}{f_{p'}(\omega)^2}.
\end{align}
This quantity is not the usual pump-probe signal but contains different phase contributions (see main text, paragraph `Differential transmission').

\section{Analytical solution of pump-probe dynamics for a 2LS}

For a two-level system, let us consider the $(2)(1)$-order equations for the cavity field and the density matrix (second order in pump, first in probe), required for the computation of the DT derived in Sec.~\ref{sec:difftransmission} (assuming a matter initial state $\rho(0)=\ket{g}\bra{g}$)
\begin{subequations}
\begin{align}
\dot\alpha^{(2)(1)}(t)&=-\left(\frac{\kappa}{2}+\mi\omega_c\right)\alpha^{(2)(1)}-\mi g \mathcal{N}\rho_{21}^{(2)(1)},\\
\dot{\rho}^{(2)(1)}_{21}(t)&=
-\left(\frac{\gamma+\gamma_\phi}{2}+\mi\omega_0\right)\rho_{21}^{(2)(1)}-\mi g \alpha^{(2)(1)}
+\mi g \alpha^{(0)(1)}(\rho_{22}^{(2)(0)}-\rho_{11}^{(2)(0)})\\\nonumber
&+\mi g \alpha^{(1)(0)}(\rho_{22}^{(1)(1)}-\rho_{11}^{(1)(1)}),
\end{align}
\end{subequations}
where we considered dissipation affecting the molecules in the form of dephasing ($\gamma_\phi$) and spontaneous emission ($\gamma$). The third-order coherence $\rho_{21}^{(2)(1)}$ is driven by the population created by the pump $(2)(0)$ as well as by the population created by both pump and probe $(1)(1)$. We rewrite the equation for the third-order coherence as
\begin{align}
\dot\rho^{(2)(1)}_{21, \tau}(t)=
-\left(\frac{\gamma+\gamma_\phi}{2}+\mi\omega_0\right)\rho_{21, \tau}^{(2)(1)}-\mi g \alpha^{(2)(1)}_\tau
+F_\tau (t),
\end{align}
where we condensed the input affecting the third-order coherence into a single term $F_\tau(t)$ and by the index $\tau$ we indicate that the quantities depend on the delay time between the pulses $\tau_\Delta$. 

\begin{figure}[t]
    \centering
    \includegraphics[width=0.8\textwidth]{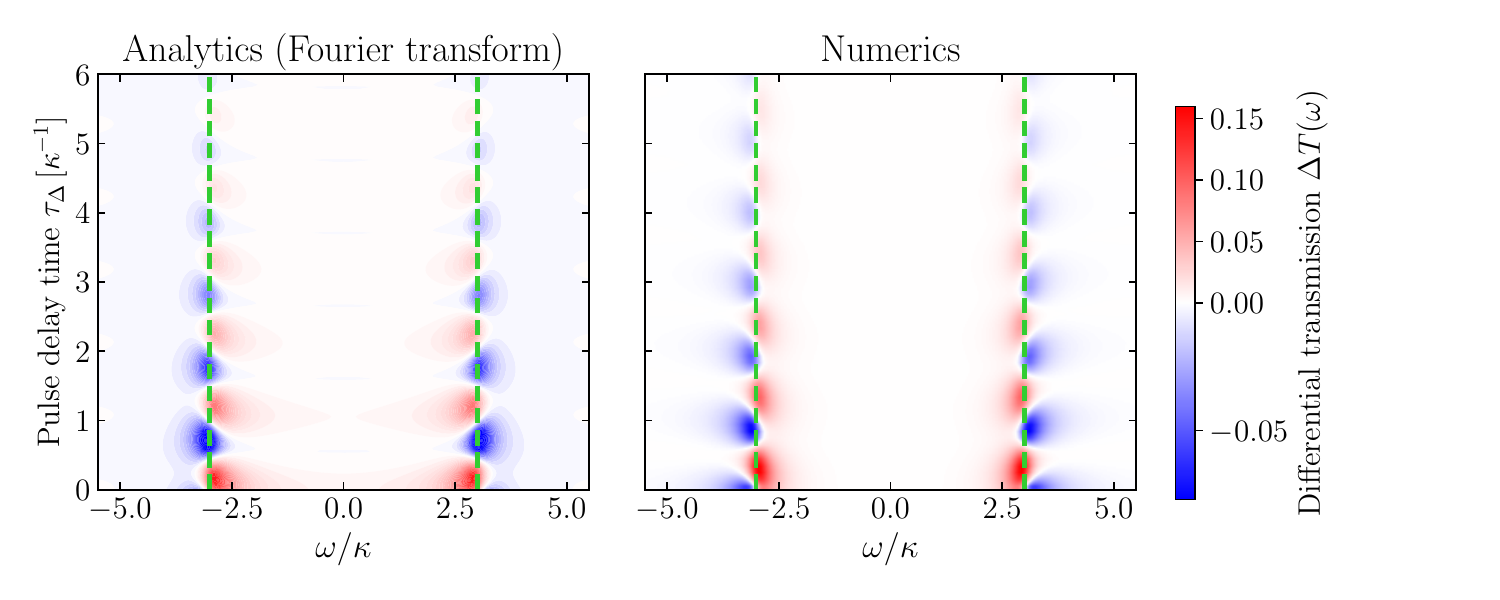}
    \caption{  Comparison between analytical (left, Fourier transform) and numerical (right, obtained by solving the differential equations) results for DT spectrum of a 2LS, plotted in a frame rotating at the central pulse frequency. The parameters are $g\sqrt{\mathcal N}=3\kappa$, $\tau_w=0.05\kappa^{-1}$, $\omega_p=\omega_{p'}=\omega_0$, and $\gamma_\phi,\gamma=0$.}
    \label{figS0}
\end{figure}

In Fourier space, we can therefore obtain the solution for $\alpha^{(2)(1)}_\tau(\omega)$
\begin{align}
\dot\alpha^{(2)(1)}_\tau(\omega)=\frac{-\mi g \mathcal{N} F_\tau (\omega)}{\left[\mi (\omega_c-\omega)+\frac{\kappa}{2}\right]\left[\mi (\omega_0-\omega)+\frac{\gamma+\gamma_\Phi}{2}\right]+g^2\mathcal{N}}.
\end{align}
From the equations of motion, we can relate second-order quantities to first-order quantities in frequency space as ($\ast$ denotes the convolution operation)
\begin{subequations}
\begin{align}
\rho_{22}^{(2)(0)}(\omega)-\rho_{11}^{(2)(0)}(\omega)&=\frac{4\mi g }{\omega+\mi\gamma}\mathrm{Im}\left[\left(\alpha^{(1)(0)}\ast \rho_{12}^{(1)(0)}\right)(\omega)\right],\\
\rho_{22}^{(1)(1)}(\omega)-\rho_{11}^{(1)(1)}(\omega)&=\frac{4\mi g }{\omega+\mi\gamma}\left\{\mathrm{Im}\left[\left(\alpha^{(0)(1)}\ast \rho_{12}^{(1)(0)}\right)(\omega)\right]+\mathrm{Im}\left[\left(\alpha^{(1)(0)}\ast \rho_{12}^{(0)(1)}\right)(\omega)\right]\right\},
\end{align}
\end{subequations}
and therefore, we can express $F_\tau (\omega)$ as a sum of double convolutions
\begin{align}
\nonumber F_\tau (\omega)=&-\frac{4g^2}{\omega+\mi\gamma}\Bigl\{\left[\alpha^{(0)(1)}\ast \mathrm{Im}\left[\alpha^{(1)(0)}\ast\rho_{12}^{(1)(0)}\right]\right]_\tau(\omega)\!+\!\left[\alpha^{(1)(0)}\ast \mathrm{Im}\left[\alpha^{(0)(1)}\ast\rho_{12}^{(1)(0)}\right]\right]_\tau(\omega)\!\\
&+\!\left[\alpha^{(1)(0)}\ast \mathrm{Im}\left[\alpha^{(1)(0)}\ast\rho_{12}^{(0)(1)}\right]\right]_\tau(\omega)\Bigr\},
\end{align}
in terms of the first-order quantities in the pump \revise{field}
\begin{align}
\alpha^{(1)(0)}(\omega)=\frac{-\tilde f_p (\omega-\omega_p)}{\left[\frac{\kappa}{2}+\mi(\omega_c-\omega)\right]+\frac{g^2\mathcal{N}}{\frac{\gamma+\gamma_\phi}{2}+\mi(\omega_0-\omega)}},\qquad \rho_{21}^{(1)(0)}(\omega)=\frac{-\mi g \alpha^{(1)(0)}(\omega)}{\mi (\omega_0-\omega)+\frac{\gamma+\gamma_\phi}{2}},
\end{align}
and similarly for the terms linear in the probe. The Fourier transform of the Gaussian pulse shapes is given by
\begin{align}
\tilde f_j (\omega)=\frac{1}{\sqrt{2\pi}}\int_{-\infty}^\infty \td t\, \me^{-\frac{(t-\tau_j)^2}{2\tau_w^2}}\me^{\mi\omega t}=\tau_w\me^{-\omega^2\tau_w^2/2}\me^{\mi\omega\tau_j},
\end{align}
where the pulse frequency and arrival time lead to displacement and an additional complex phase in frequency space, respectively. A comparison between the Fourier transform result and the numerical result obtained by solving the differential equations in time domain is shown in Fig.~\ref{figS0}.

\section{Formal integration of pump-probe response}

From Eq.~\eqref{eq:pumpprobedensity}, the integrated solution for the vectorized density matrix second order in the pump and first order in the probe,  $\vec{\rho}^{(2)(1)}$, is obtained as [denoting the free molecular propagator, i.e., the Green's function by $\theta (t-t')\me^{-\mi\mathcal{L}_0 (t-t')}\equiv\mathcal{G}(t-t')$]
\begin{align}
\label{eq:timev}
&\vec{\rho}^{(2)(1)}(t)=-\mi\int_0^t \td t_1\, \mathcal{G}(t-t_1)\times\\\nonumber
&\times\left[\mathcal{L}\ts{int}^{(2)(1)}(t_1)\vec{\rho}^{(0)(0)}(t_1)+\mathcal{L}\ts{int}^{(2)(0)}(t_1)\vec{\rho}^{(0)(1)}(t_1)\!+\!\mathcal{L}\ts{int}^{(1)(1)}(t_1)\vec{\rho}^{(1)(0)}(t_1)\!+\!\mathcal{L}\ts{int}^{(1)(0)}(t_1)\vec{\rho}^{(1)(1)}(t_1)\!+\!\mathcal{L}\ts{int}^{(0)(1)}(t_1)\vec{\rho}^{(2)(0)}(t_1)\right].
\end{align}
Writing out all terms explicitly in terms of the molecular initial state $\vec\rho (0)$, yields the following contributions: 
\begin{align}
\label{eq:pumpprobeintegrals}
\nonumber\vec{\rho}^{(2)(1)}(t)&=-\mi\int_0^t \td t_1\,\mathcal{G}(t-t_1)\mathcal{L}\ts{int}^{(2)(1)}(t_1)\vec{\rho}(0)\\\nonumber
&(-\mi)^2\int_0^t \td t_2\,\int_0^{t_2}\td t_1\,\mathcal{G}(t-t_2)\mathcal{L}\ts{int}^{(1)(1)}(t_2) \mathcal{G}(t_2-t_1)\mathcal{L}\ts{int}^{(1)(0)}(t_1)\vec\rho(0)\\\nonumber
&(-\mi)^2\int_0^t \td t_2\,\int_0^{t_2}\td t_1\,\mathcal{G}(t-t_2)\mathcal{L}\ts{int}^{(2)(0)}(t_2)\mathcal{G}(t_2-t_1)\mathcal{L}\ts{int}^{(0)(1)}(t_1)\vec\rho (0)\\\nonumber
&(-\mi)^2\int_0^t \td t_2\,\int_0^{t_2} \td t_1\,\mathcal{G}(t-t_2)\mathcal{L}\ts{int}^{(0)(1)}(t_2)\mathcal{G}(t_2-t_1)\mathcal{L}\ts{int}^{(2)(0)}(t_1)\vec\rho (0)\\\nonumber
&(-\mi)^2\int_0^t \td t_2\,\int_0^{t_2} \td t_1\,\mathcal{G}(t-t_2)\mathcal{L}\ts{int}^{(1)(0)}(t_2)\mathcal{G}(t_2-t_1)\mathcal{L}\ts{int}^{(1)(1)}(t_1)\vec\rho (0)\\\nonumber
&(-\mi)^3\int_0^t \td t_3\,\int_0^{t_3} \td t_2\,\int_0^{t_2} \td t_1\,\mathcal{G}(t-t_3)\mathcal{L}\ts{int}^{(1)(0)}(t_3)\mathcal{G}(t_3-t_2)\mathcal{L}\ts{int}^{(1)(0)}(t_2)\mathcal{G}(t_2-t_1)\mathcal{L}\ts{int}^{(0)(1)}(t_1)\vec\rho (0)\\\nonumber
&(-\mi)^3\int_0^t \td t_3\,\int_0^{t_3} \td t_2\,\int_0^{t_2} \td t_1\,\mathcal{G}(t-t_3)\mathcal{L}\ts{int}^{(1)(0)}(t_3)\mathcal{G}(t_3-t_2)\mathcal{L}\ts{int}^{(0)(1)}(t_2)\mathcal{G}(t_2-t_1)\mathcal{L}\ts{int}^{(1)(0)}(t_1)\vec\rho (0)\\
&(-\mi)^3\int_0^t \td t_3\,\int_0^{t_3} \td t_2 \,\int_0^{t_2} \td t_1\,\mathcal{G}(t-t_3)\mathcal{L}\ts{int}^{(0)(1)}(t_3)\mathcal{G}(t_3-t_2)\mathcal{L}\ts{int}^{(1)(0)}(t_2)\mathcal{G}(t_2-t_1)\mathcal{L}\ts{int}^{(1)(0)}(t_1)\vec\rho (0).
\end{align}
The last term corresponds to the result that would be obtained for conventional free-space spectroscopy with well-separated pulses, i.e., two linear applications of the pump field followed by application of the probe field~\cite{mukamel1995principles}.

\begin{figure}[t]
    \centering
    \includegraphics[width=0.92\textwidth]{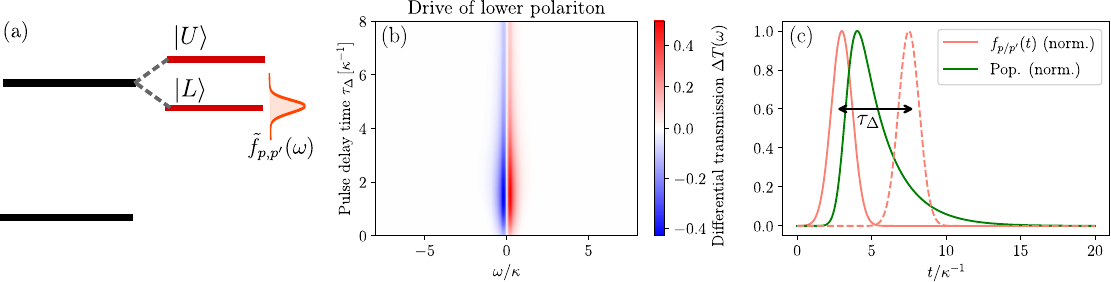}
    \caption{  (a) The envelope of the pump and probe pulses in frequency domain is fitted to the lower polariton frequency. (b) DT spectrum for excitation of the lower polariton. We chose $g\sqrt{\mathcal{N}}=3\kappa$ and a pulse width of $\tau_w=\kappa^{-1}$. Note that, since we are in a rotating frame, $\omega=0$ corresponds to the central pulse (and therefore the lower polariton) frequency. (c) Time evolution of molecular population created by the pump.}
    \label{figS1}
\end{figure}

\section{Narrowband spectroscopy of polaritons}

The results presented in the main text were obtained by broadband excitation of the polaritons, i.e., by exciting an (equal) superposition of upper and lower polariton. Instead, one may also consider narrowband excitation of the polaritons, by choosing a pulse width in frequency domain $\tau_w^{-1}< g\sqrt{\mathcal{N}}$ \revise{and fitting the pulse carrier frequency to the polariton frequencies}. Figs.~\ref{figS1}(b), (c) show the result obtained for narrowband pump-probe excitation of the lower polariton. In this case, no Rabi oscillations are observed since no superposition of upper and lower polariton is excited. Narrowband excitation of the upper polariton shows a similar result.

\begin{figure}[t]
    \centering
    \includegraphics[width=0.99\textwidth]{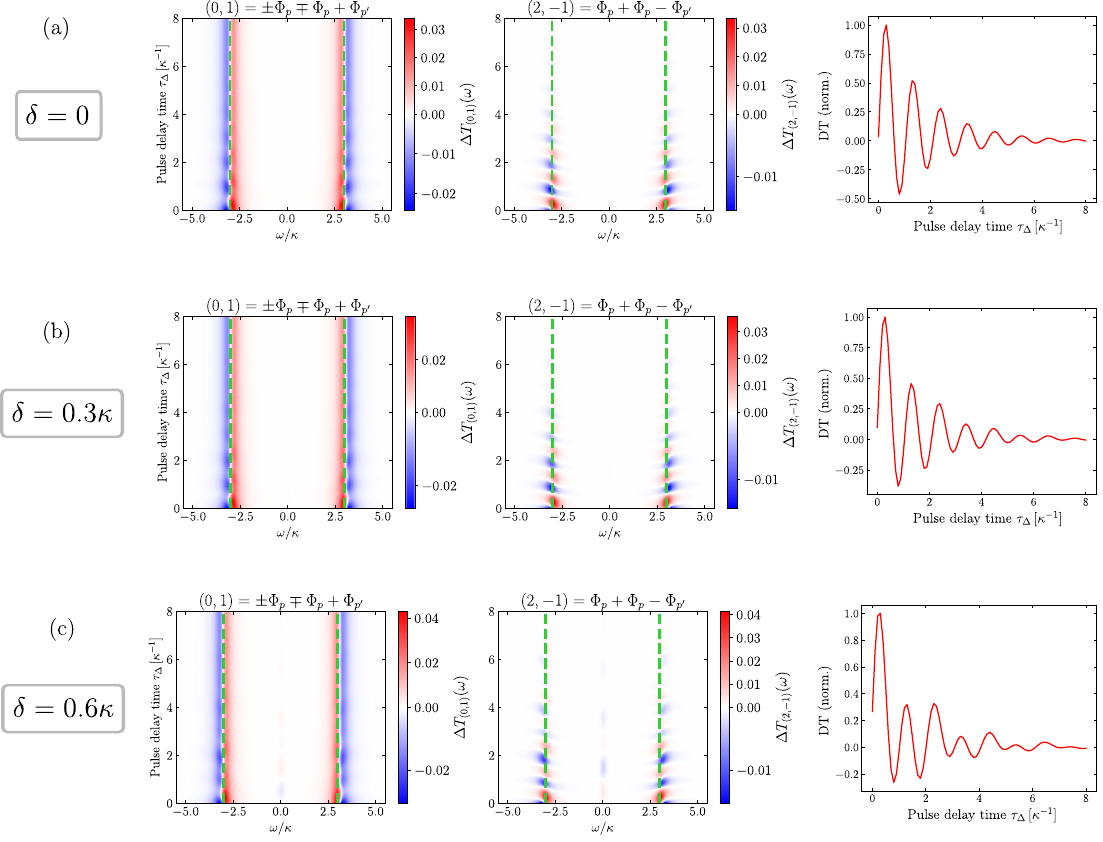}
    \caption{\revise{DT for varying degrees of disorder (a) $\delta=0$, (b) $\delta=0.3\kappa$, and (c) $\delta=0.6\kappa$. Left and middle columns show different phase contributions, while the right column presents a cut through the $(2,-1)$ contribution at $\omega=g\sqrt{\mathcal{N}}$ (normalized for comparison across different cases). We chose $g\sqrt{\mathcal{N}}=3\kappa$, $\gamma_\phi=0.2\kappa$.}}
    \label{figS2}
\end{figure}

\section{Effects of disorder on DT}

\revise{To study the effect of inhomogeneous broadening, we employ a simplified model in which the molecular ensemble is divided into two (equal) subensembles with shifted transition frequencies, $\omega_{\pm}=\omega_0\pm\delta$. In Fig.~\ref{figS2}, we illustrate the effect of the degree of disorder $\delta$ on the DT spectra. Generally, we find that the DT appears quite robust with increasing disorder. At larger degrees of disorder, the dark states are appearing in the spectrum at $\omega=0$, and there is a slight increase in the vacuum Rabi splitting, consistent with previous theoretical treatments \cite{sommer2021molecular, schwennicke2024extracting, chen2022interplay}. Furthermore, the bright state lifetime [as extracted from the DQC $(2,-1)$ contribution] decays faster with increasing disorder.}

\section{Bright and dark states in mean-field theory}

\revise{We show here how molecular dark and bright states are arising in the mean-field approach (for a 2LS) \cite{fowler2022efficient}. The total population is given by
\begin{align}
p_\mathrm{tot}= \sum_{j=1}^\mathcal{N}\expval{\s_j^\dagger\s_j},
\end{align}
which in mean field theory becomes
\begin{align}
p_\mathrm{tot}=\mathcal{N}\expval{\s^\dagger\s}.
\end{align}
Using that the general population of the $k$-th collective mode is given by $p_{k}=\expval{\s_k^\dagger\s_k}$, the population of the bright mode (corresponding to $k=0$) is given by
\begin{align}
p_B=p_{k=0}=\frac{1}{\mathcal{N}}\sum_{j,j'=1}^\mathcal{N}\expval{\s_j^\dagger\s_{j'}^{\phantom{\dagger}}}=\frac{1}{\mathcal{N}}\sum_{j=1}^\mathcal{N}\expval{\s_j^\dagger\s_j^{\phantom{\dagger}}}+\frac{1}{\mathcal{N}}\sum_{j\neq j'}\expval{\s_j}^*\expval{\s_{j'}},
\end{align}
which in mean field theory becomes
\begin{align}
p_B=\expval{\s^\dagger\s}+(\mathcal{N}-1)|\expval{\s}|^2
\end{align}
The dark state population $p_D=p_\mathrm{tot}-p_B$ is therefore given 
\begin{align}
p_D=(\mathcal{N}-1)\expval{\s^\dagger\s}-(\mathcal{N}-1)|\expval{\s}|^2.
\end{align}
In the large $\mathcal{N}$ limit, this can be expressed as
\begin{subequations}
\begin{align}
p_B&\approx \expval{\s^\dagger\s}+\mathcal{N}|\expval{\s}|^2,\\
p_D&\approx\mathcal{N}\left(\expval{\s^\dagger\s}-|\expval{\s}|^2\right).
\end{align}
\end{subequations}
Therefore, the bright state corresponds to molecular coherence in the large $\mathcal{N}$ limit.}

\section{Details on phase expansion}

For an arbitrary multilevel system described by a Hamiltonian $\mathcal H_0$ and a dipole operator $\hat\mu=\hat{\mu}^{(+)}+\hat{\mu}^{(-)}$, the mean-field Hamiltonian in the rotating wave approximation is given by
\begin{align}
\mathcal{H}\ts{MF}=\mathcal{H}_0+ E_0\left(\alpha^*(t)\hat\mu^{(-)}+\alpha (t)\hat\mu^{(+)}\right).
\end{align}
We expand the cavity field and density matrix \revise{in terms of the input fields' amplitudes and phases} as
\begin{align}
\alpha (t)=\sum_{n,m, \vec{v}}\eta_p^n \eta_{p'}^m \me^{-\mi\vec{v}\cdot \vec{\Phi}} \alpha^{(n)(m)}_{\vec{v}} (t),\quad \rho (t)=\sum_{n,m, \vec{v}}\eta_p^n \eta_{p'}^m \me^{-\mi\vec{v}\cdot \vec{\Phi}}\rho^{(n)(m)}_{\vec{v}}(t),
\end{align}
where  $\vec\Phi =(\Phi_p,\Phi_{p'})^\top$ denotes the vector of pump and probe phases and $\vec v=(v_p, v_{p'})^\top$ denotes the coefficients of the phases, respectively. Possible values for the phase coefficients in $(n), (m)$-th order are \revise{$v_p (n)=\sum_{j=1}^n s_j$, $v_{p'}(m)=\sum_{j=1}^m s_j$, where each $s_j=\pm 1$}. Note that complex conjugation corresponds to flipping the sign of the phase, i.e., $\left[\alpha^{(n)(m)}_{\vec v}\right]^*=\alpha^{(n)(m)*}_{-\vec v}$.

The equation for the $(n)(m)$-th order cavity field including phases on the pump and probe input fields reads
\begin{align}
\dot{\alpha}^{(n)(m)}(t) &=-\left(\frac{\kappa}{2}+\mi\omega_c\right)\alpha^{(n)(m)}-\mi E_0\mathcal{N}\mathrm{Tr}[\hat{\mu}^{(-)}\rho^{(n)(m)}] 
-\delta_{1n}\delta_{0m} f_p (t-\tau_p)\me^{-\mi\omega_p t}\me^{-\mi \Phi_p} 
-\delta_{0n}\delta_{1m}f_{p'}(t-\tau_{p'})\me^{-\mi\omega_{p'} t}\me^{-\mi\Phi_{p'}}.
\end{align}
It is now straightforward to see that, at first order, the only phase contributions of the cavity field which are driven by the input pulses are $(1,0)$ for the field created by the pump and $(0,1)$ for the field created by the probe
\begin{subequations}
\begin{align}
\dot{\alpha}^{(1)(0)}_{(1,0)}&=-\left(\frac{\kappa}{2}+\mi\omega_c\right)\alpha^{(1)(0)}_{(1,0)} 
-\mi E_0\mathcal{N}\mathrm{Tr}[\hat\mu^{(-)}\rho^{(1)(0)}_{(1,0)}] - f_p(t-\tau_p)\me^{-\mi\omega_p t}, \\
\dot{\alpha}^{(0)(1)}_{(0,1)}&=-\left(\frac{\kappa}{2}+\mi\omega_c\right)\alpha^{(0)(1)}_{(0,1)} 
-\mi E_0\mathcal{N}\mathrm{Tr}[\hat\mu^{(-)}\rho^{(1)(0)}_{(1,0)}] - f_{p'}(t-\tau_{p'})\me^{-\mi\omega_{p'} t},
\end{align}
\end{subequations}
while the other possible phase combinations at first order $(-1,0)$ and $(0,-1)$ are contained in  the complex conjugate cavity fields. All other phase components are not driven at first order and are therefore zero. The positive-phase cavity fields can create positive-phase density matrices in first order via the application of $\hat\mu^{(+)}$ to a ket (creation of excitation) or to a bra (annihilation of excitation):
\begin{subequations}
\begin{align}
\dot{\rho}^{(1)(0)}_{(1,0)}&=
-\mi[\mathcal{H}_0, \rho^{(1)(0)}_{(1,0)}] 
+\mathcal{D}[\rho^{(1)(0)}_{(1,0)}]-\mi E_0 \alpha^{(1)(0)}_{(1,0)}[\hat\mu^{(+)}, \rho(0)], \\
\dot{\rho}^{(0)(1)}_{(0,1)}&=
-\mi[\mathcal{H}_0, \rho^{(0)(1)}_{(0,1)}] +\mathcal{D}[\rho^{(0)(1)}_{(0,1)}]
-\mi E_0 \alpha^{(0)(1)}_{(0,1)}[\hat\mu^{(+)}, \rho(0)],
\end{align}
\end{subequations}
while the complex conjugate processes correspond to the application of $\hat\mu^{(-)}$, thereby imprinting negative phase contributions onto the density matrix. 

Beyond first order, the cavity field is not driven directly anymore. Instead, the density matrix is now driven by lower orders of $\alpha$, which determines the possible phase contributions in higher orders. The phase combinations for the $(n)(m)$-th order density matrix driven by lower orders of cavity field and density matrix are given by
\begin{align}
\dot{\rho}^{(n)(m)}_{\vec{v}}=-\mi[\mathcal{H}_0,\rho^{(n)(m)}_{\vec{v}}]+\mathcal{D}[\rho^{(n)(m)}_{\vec{v}}]-\mi E_0\sum_{j=0}^{n-1}\sum_{j'=0}^{m-1}\sum_{ \vec{u}+\vec{w}=\vec{v}}\left(\alpha_{\vec{u}}^{(n-j)(m-j')}[\hat\mu^{(+)},\rho^{(j)(j')}_{\vec{w}}]+\alpha_{\vec{u}}^{(n-j)(m-j')*}[\hat\mu^{(-)},\rho^{(j)(j')}_{\vec{w}}]\right),
\end{align}
where the possible values for $\vec{u}$, $\vec{w}$ are determined by the phase contributions in lower orders. This drives the phase contributions for the cavity field at nonlinear orders \revise{$n+m> 1$} as
\begin{align}
\dot{\alpha}^{(n)(m)}_{\vec v}(t) =-\left(\frac{\kappa}{2}+\mi\omega_c\right)\alpha^{(n)(m)}_{\vec v}-\mi E_0\mathcal{N}\mathrm{Tr}[\hat{\mu}^{(-)}\rho^{(n)(m)}_{\vec v}].
\end{align}

\section{Explicit phase expansion for three-level systems}

Let us explicitly consider the phase expansion for a three-level system (3LS), which can, e.g., serve as a simplified model for an anharmonic vibrational mode, described by the dipole operator
\begin{align}
\hat\mu=\mu_{12}\hat{\s}_{12}+\mu_{23}\hat{\s}_{23}+\mathrm{H.c.},
\end{align}
where $\hat{\s}_{12}=\ket{1}\bra{2}$ and $\hat{\s}_{23}=\ket{2}\bra{3}$ are the corresponding lowering operators. The mean-field Hamiltonian of a 3LS then reads
\begin{align}
\mathcal{H}_\text{MF} &=\omega_2\hat{\s}_{12}^\dagger\hat{\s}_{12}^{\phantom{\dagger}}+\omega_3\hat{\s}_{23}^\dagger\hat{\s}_{23}^{\phantom{\dagger}}+ g_{12}\left(\hat{\s}_{12}^\dagger\alpha (t)+\mathrm{h.c.}\right)+g_{23}\left(\hat{\s}_{23}^\dagger\alpha (t)+\mathrm{h.c.}\right),
\end{align}
with the transition frequencies $\omega_2$, $\omega_3$ (we set $\omega_1=0$), and $g_{12}=\mu_{12}E_0$, $g_{23}=\mu_{23}E_0$ describe the coupling strengths of the $\ket{1}$-$\ket{2}$ and $\ket{2}$-$\ket{3}$ transitions \revise{to the cavity field}, respectively. The 2LS limit can simply be obtained by setting $g_{23}=0$. In addition, we will consider dephasing affecting the transitions which can be described by the dissipator
\begin{align}
\mathcal{D}[\rho]=\gamma_\phi\left[\hat{\s}_{12}^\dagger\hat{\s}_{12}^{\phantom{\dagger}}\rho\hat{\s}_{12}^\dagger\hat{\s}_{12}-\frac{1}{2}\left\{\rho,\hat{\s}_{12}^\dagger\hat{\s}_{12}^{\phantom{\dagger}}\right\}\right]+\gamma_\phi\left[\hat{\s}_{23}^\dagger\hat{\s}_{23}^{\phantom{\dagger}}\rho\hat{\s}_{23}^\dagger\hat{\s}_{23}^{\phantom{\dagger}}-\frac{1}{2}\left\{\rho,\hat{\s}_{23}^\dagger\hat{\s}_{23}^{\phantom{\dagger}}\right\}\right],
\end{align}
where $\{\cdot,\cdot\}$ denotes the anticommutator.\\

\paragraph{First order.} Assuming identical frequencies for pump and probe $\omega_{p}=\omega_{p'}$ and an initial state $\rho(0)=\ket{g}\bra{g}$, the equations for the first order coherences in a frame rotating at $\omega_p$ are given by
\begin{subequations}
\begin{align}
\dot{\rho}_{21,(1,0)}^{(1)(0)}&=-\left(\frac{\gamma_\phi}{2}-\mi\Delta_{12}\right){\rho}_{21,(1,0)}^{(1)(0)}-\mi g_{12} \alpha^{(1)(0)}_{(1,0)},\\
\dot{\rho}^{(0)(1)}_{21, (0,1)}&=-\left(\frac{\gamma_\phi}{2}-\mi\Delta_{12}\right)\rho^{(0)(1)}_{21, (0,1)}-\mi g_{12} \alpha^{(0)(1)}_{(0,1)},
\end{align}
\end{subequations}
while the equation for the first-order fields are given by
\begin{subequations}
\begin{align}
\dot{\alpha}_{(1,0)}^{(1)(0)}&= -\left(\frac{\kappa}{2}-\mi\Delta_c\right)\alpha^{(1)(0)}_{(1,0)}-\mi \mathcal{N} g_{12} \rho_{21, (1,0)}^{(1)(0)}- f_p (t-\tau_p),\\
\dot{\alpha}^{(0)(1)}_{(0,1)}&=-\left(\frac{\kappa}{2}-\mi\Delta_c\right)\alpha_{(0,1)}^{(0)(1)}-\mi\mathcal{N} g_{12} \rho_{21, (0,1)}^{(0)(1)}-f_{p'}(t-\tau_{p'}),
\end{align}
\end{subequations}
where we define the detunings w.r.t.~the central input pulse frequency $\Delta_c=\omega_p-\omega_c$, $\Delta_{12}=\omega_p-\omega_2$, and $\Delta_{23}=\omega_p-(\omega_3-\omega_2)$. The other phase combinations $(-1,0)$ and $(0,-1)$ are contained in the complex conjugate cavity fields.

\paragraph{Second order.} The second order populations and coherences are given by 
\reviseTwo{\begin{subequations}
\begin{align}
\dot{\rho}_{22,(0,0)}^{(2)(0)}&=-\mi g_{12} \alpha_{(1,0)}^{(1)(0)}\rho_{12, (-1,0)}^{(1)(0)}+\mi g_{12} \alpha_{(-1,0)}^{(1)(0)*}\rho_{21, (1,0)}^{(1)(0)},\\
\dot{\rho}_{22, (1,-1)}^{(1)(1)}&=-\mi g_{12} \alpha_{(1,0)}^{(1)(0)}\rho_{12, (0,-1)}^{(0)(1)}+\mi g_{12} \alpha_{(0,-1)}^{(0)(1)*}\rho_{21, (1,0)}^{(1)(0)},\\
\dot{\rho}_{31, (2,0)}^{(2)(0)}&=-\left[\frac{\gamma_\phi}{2}-\mi(\Delta_{12}+\Delta_{23})\right]\rho_{31,(2,0)}^{(2)(0)}-\mi g_{23}\alpha_{(1,0)}^{(1)(0)}\rho_{21,(1,0)}^{(1)(0)},\\
\dot{\rho}_{31, (1,1)}^{(1)(1)}&=-\left[\frac{\gamma_\phi}{2}-\mi(\Delta_{12}+\Delta_{23})\right]\rho_{31,(1,1)}^{(1)(1)}-\mi g_{23}\alpha_{(0,1)}^{(0)(1)}\rho_{21,(1,0)}^{(1)(0)}-\mi g_{23}\alpha^{(1)(0)}_{(1,0)}\rho_{21,(0,1)}^{(0)(1)},
\end{align}
\end{subequations}}
while all second-order cavity fields are not driven and therefore remain zero at all times. The other phase combinations in second order $(-1,1)$, $(-2,0)$, and $(-1,-1)$ are obtained as the complex conjugates of the above equations.

\begin{figure}[t]
    \centering
    \includegraphics[width=0.99\textwidth]{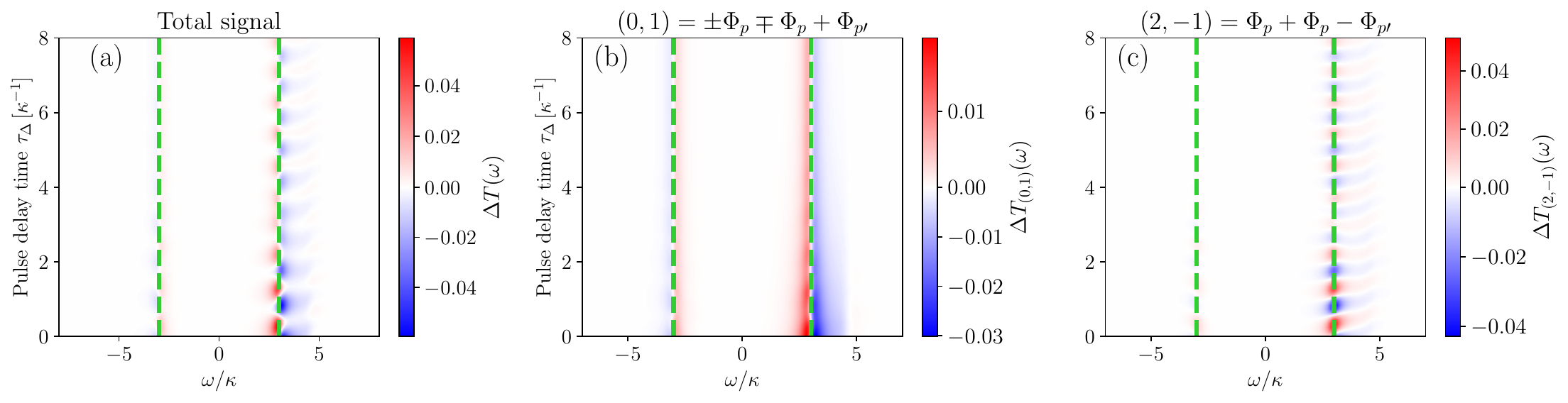}
    \caption{(a) Total DT signal as a function of pulse delay time for a 3LS, as well as phase contributions (b) $(0,1)$, and (c) $(2,-1)$. We chose $g_{12}\sqrt{\mathcal N}=3\kappa$, $g_{23}=\sqrt{2}g_{12}$, $\omega_c=\omega_p=\omega_{p'}=\omega_{2}$, and $\omega_3-\omega_2=1.8\omega_2$. The dephasing was set to $\gamma_\phi=0.1\kappa$.}
    \label{figS3}
\end{figure}

\paragraph{Third order.} The third-order coherences are given by
\reviseTwo{\begin{subequations}
\begin{align}
\dot{\rho}_{21,(0,1)}^{(2)(1)}&=-\left(\frac{\gamma_\phi}{2}-\mi\Delta_{12}\right){\rho}_{21,(0,1)}^{(2)(1)}-\mi g_{12} \alpha^{(2)(1)}_{(0,1)}+2\mi g_{12} \alpha_{(0,1)}^{(0)(1)}\rho_{22, (0,0)}^{(2)(0)}+2\mi g_{12} \alpha_{(1,0)}^{(1)(0)}\rho_{22, (-1,1)}^{(1)(1)}-\mi g_{23}\alpha^{(1)(0)*}_{(-1,0)}\rho_{31, (1,1)}^{(1)(1)},\\
\dot{\rho}^{(2)(1)}_{21, (2,-1)}&=-\left(\frac{\gamma_\phi}{2}-\mi\Delta_{12}\right)\rho^{(2)(1)}_{21, (2,-1)}-\mi g_{12} \alpha^{(2)(1)}_{(2,-1)}+2\mi g_{12} \alpha_{(1,0)}^{(1)(0)}\rho_{22, (1,-1)}^{(1)(1)}-\mi g_{23}\alpha^{(0)(1)*}_{(0,-1)}\rho_{31, (2,0)}^{(2)(0)},\\
\dot{\rho}^{(2)(1)}_{32, (0,1)}&=-\left(\gamma_\phi-\mi \Delta_{23} \right)\rho_{32, (0,1)}^{(2)(1)}-\mi g_{23}\alpha^{(0)(1)}_{(0,1)}\rho^{(2)(0)}_{22, (0,0)}-\mi g_{23}\alpha^{(1)(0)}_{(1,0)}\rho^{(1)(1)}_{22, (-1,1)}+\mi g_{12}\alpha^{(1)(0)*}_{(-1,0)}\rho_{31, (1,1)}^{(1)(1)}, \\
\dot{\rho}^{(2)(1)}_{32, (2,-1)}&=-\left(\gamma_\phi-\mi\Delta_{23}\right)\rho_{32, (2,-1)}^{(2)(1)}-\mi g_{23}\alpha^{(1)(0)}_{(1,0)}\rho_{22, (1,-1)}^{(1)(1)}+\mi g_{12}\alpha^{(0)(1)*}_{(0,-1)}\rho_{31, (2,0)}^{(2)(0)}.
\end{align}
\end{subequations}}
The third-order equations for the fields are given by
\begin{subequations}
\begin{align}
\dot{\alpha}_{(0,1)}^{(2)(1)}&=-\left(\frac{\kappa}{2}-\mi\Delta_c\right)\alpha_{(0,1)}^{(2)(1)}-\mi\mathcal{N} g_{12} \rho_{21, (0,1)}^{(2)(1)}-\mi\mathcal{N} g_{23}\rho_{32, (0,1)}^{(2)(1)},\\
\dot{\alpha}^{(2)(1)}_{(2,-1)}&=-\left(\frac{\kappa}{2}-\mi\Delta_c\right)\alpha_{(2,-1)}^{(2)(1)}-\mi\mathcal{N} g_{12} \rho_{21, (2,-1)}^{(2)(1)}-\mi\mathcal{N} g_{23} \rho_{32, (2,-1)}^{(2)(1)}.
\end{align}
\end{subequations}
The other phase combinations $(0,-1)$, and $(-2,1)$ are described by the complex conjugate cavity fields. The resulting DT spectrum for the different phase contributions is plotted in Fig.~\ref{figS3}, showing that for a 3LS the $(2,-1)$-contribution also exists for pulse delay times $>(\kappa+\gamma_\phi)^{-1}$.

\section{Extension to multimode cavities}

\revise{For many scenarios and applications (e.g., studies of polariton transport, polariton condensation,...), the picture of a single cavity mode does not suffice and instead many photonic modes inside the cavity are required. We sketch here how the mean-field treatment can be extended in a straightforward way to also account for such more complicated electromagnetic environments, namely by grouping the molecular system into subsystems \cite{fowlerwrightthesis}. 

To this end, we consider a planar cavity which extends a width $W$ perpendicular (transverse) to the cavity axis. To enable a tractable numerical treatment, one can coarse-grain the transverse extent into $N_k$ discrete points at distance $\Delta r$, $W=N_k\Delta r$. Assuming the total number of molecules $\mathcal{N}$  inside the cavity to be uniformly distributed along the transverse direction, every one of the grid points is assigned $N_E=\mathcal{N}/N_k$ molecules interacting with $N_k$ photon modes. This is described by the Hamiltonian
\begin{align}
\mathcal{H} = \sum_{k=1}^{N_k}\omega_k \hat{a}_k^\dagger \hat{a}_k^{\phantom{\dagger}} + \sum_{n=1}^{N_k}\sum_{j=1}^{N_E}\left[\mathcal{H}_0^{nj}+E_0\sum_{k=1}^{N_k}\left(\hat{a}_k\hat{\mu}_{nj}^{(+)}\me^{\mi k r_n}+\mathrm{h.c.}\right)\right],
\end{align}
where the index $n=1,\hdots, N_k$ labels the position while $j=1,\hdots N_E$ labels the particular molecule at a certain position. For the cavity, one may assume a quadratic dispersion $\omega_k=\omega_c+k^2 c^2/(2\omega_c)$, with $\omega_c$ the cavity frequency at $k=0$. This parabolic approximation is well‐known to accurately capture the mode structure of planar cavities near normal incidence. Within the mean-field approximation, the cavity field is assumed to be classical and all molecules at a position $r_n$ are assumed to be identical, i.e., $\mathcal{H}_0^{nj}=\mathcal{H}_0^{n}$, $\hat{\mu}_{nj}=\hat{\mu}_n$, leading to the simplified Hamiltonian
\begin{align}
\mathcal{H}_\mathrm{MF}=\sum_{n=1}^{N_k}\mathcal{H}_{0}^n+E_0\sum_{k=1}^{N_k}\sum_{n=1}^{N_k}\left(\alpha_k (t) \hat{\mu}_n^{(+)}\me^{\mi k r_n}+\mathrm{h.c.}\right).
\end{align}
The classical evolution of the $N_k$ cavity modes driven by pump and probe input pulses with wavevectors $k_p$ and $k_{p'}$ transverse to the cavity axis is given by
\begin{align}
\dot{\alpha}_k (t) =-\left(\mi\omega_k +\frac{\kappa}{2}\right)\alpha_k (t) -\mi E_0 N_E\sum_{n=1}^{N_k} \expval{\hat{\mu}_n^{(-)}(t)}\me^{-\mi k r_n}-\sum_{j=p,p'}\eta_j \delta_{k_j,k} f_j (t-\tau_j)\me^{-\mi\omega_j t},
\end{align}
where only the modes corresponding to $k=k_p, k_{p'}$ are driven by the input pulses and we assume an equal decay rate $\kappa$ for all cavity modes. This set of equations can now, e.g., be used to model pump-probe spectroscopy with pulses entering the cavity at oblique angles.}

\end{document}